\documentclass[a4paper,12pt]{article}
\usepackage{latexsym}
\usepackage{graphicx}

\def\bnabla{\mbox{\boldmath{$\nabla$}}}
\def\bomega{\mbox{\boldmath{$\omega$}}}
\def\V{{\ensuremath\mathbf v}}

\def\D{\partial}
\def\EC{\end{center}} 
\def\BC{\begin{center}} 
\def\sfrac#1#2{\mbox{$\frac{#1}{#2}$}}
\def\BAN{\begin{eqnarray*}}
\def\EAN{\end{eqnarray*}}
\def\BA{\begin{eqnarray}}
\def\EA{\end{eqnarray}}
\def\BE{\begin{equation}}
\def\EE{\end{equation}}
\def\NN{\nonumber}

\title{Modeling transitional plane Couette flow}
\author{{\large Maher Lagha and Paul Manneville}\\
Laboratoire d'Hydrodynamique (LadHyX),\\
CNRS-\'Ecole Polytechnique, F-91128 Palaiseau, France.}
\date{Eur. Phys. J. B {\bf58} (2007) 433--447.}

\begin{document}

\maketitle

\begin{abstract}
The Galerkin method is used to derive a realistic model of
plane Couette flow in terms of partial differential equations governing
the space-time dependence of the amplitude of a few cross-stream modes.
Numerical simulations show that it reproduces the globally sub-critical
behavior typical of this flow. In particular, the statistics of turbulent
transients at decay from turbulent to laminar flow displays striking
similarities with experimental findings.
\end{abstract}
{\bf PACS}: 47.27.Cn, 47.60.+i

\section{Introduction}
\label{intro}

\sloppy

The transition to turbulence in shear flows close to a solid wall
is far from being completely understood. This situation arises because
linear stability analysis is less fruitful for systems in which
transient algebraic energy growth may become relevant than when
exponentially growing modes are present~\cite{SH01}. It seems indeed
easier to understand the case of super-critical 
instabilities (especially in closed-flow configurations such as
Rayleigh--B\'enard convection and the Taylor--Couette instability
for which the classical tools of weakly nonlinear analysis are
available) than the case of discontinuous transitions marked by
a competition between solutions derived from linear theory and other
fully nonlinear solutions. An interesting review of the main issues
related to this problem can be found in~\cite{MK05}.
A classical example of {\it globally sub-critical\/} transition
is plane Poiseuille flow ---the flow between 
two fixed parallel plates driven by a pressure gradient--- which is
linearly unstable only beyond some high Reynolds number
$R_{\rm c}$ \cite{Or71} while {\it turbulent spots}, {\it i.e.}
patches of turbulent flow scattered amidst laminar flow and separated
from it by well defined fronts,
can develop for $R$ as low as about $R_{\rm c}/4$ \cite{CWP82}. The
situation is even worse for plane Couette flow (pCf in the following),
the simple shear
flow driven by two plates moving parallel to one another, which is
stable for all $R$, i.e. $R_{\rm c}=\infty$ \cite{Ro73}, but experiences a
direct transition to turbulence at moderate values of
$R$ \cite{LJ91,TA92,DD94,PD05}. {\it Linear\/} theory, modal and
non-modal~\cite{SH01}, arguably indicates ways to escape from the
laminar regime but is of limited help to understand the main problem
alluded to above, namely the nature of the nontrivial {\it nonlinear\/}
state and the coexistence in physical space of these different
solutions.

Direct numerical simulations of the Navier--Stokes equations (DNS)
have been intensively used to identify elementary processes involved
in the sustenance of the nontrivial turbulent state,
see e.g. \cite{HKW95}. An interesting approach was the reduction to a
{\it minimal flow unit} (MFU) introduced by Jimenez and Moin
\cite{JM91}, in which the dimensions of the simulation domain were
decreased down to sizes for which turbulence is just
maintained. The resulting dynamics can then be understood as a
chaotic evolution of large coherent structures driving smaller eddies
in a more or less stochastic way (though strict separation of scales is not
legitimate). From observations of near-wall turbulent flow \cite{HKW95},
Waleffe was able to derive a differential model involving the amplitude
of a few modes associated to such coherent structures
\cite{Wa97}, in the spirit of a Galerkin approach that was made more
systematic by Eckhardt and co-workers \cite{SE97,EM99,MFE04} using
Fourier modes or by Holmes and co-workers \cite{SMH05} using empirical
modes extracted from DNS by proper orthogonal decomposition.

Such models indeed help one to illustrate some of the mechanisms
involved in the sustenance of turbulence or the competition between
laminar and turbulent states in phase space. However, they are
intrinsically and explicitly 
low-dimensional, which situates the problem within the context of
discrete dynamical systems appropriate to {\it temporal\/} chaos,
e.g. in discussing transients in terms of fractal border of the
laminar-flow basin of attraction in phase space
\cite{EF05}. Therefore, they do not allow one to approach such
problems as spot propagation that require full reference to the
physical space. When similar questions 
were posed for convection, {\it i.e.} pattern formation and
{\it spatio-temporal chaos\/}, it was found particularly valuable to
pass from ordinary differential systems that can only deal with
temporal behavior, such as the Lorenz model, to partial differential
equations that also account for spatial dependence, {\it e.g.}
the Swift--Hohenberg model \cite{SH77} and its numerous variants.
The present paper is devoted to the derivation and use
of such a closed set of partial differential equations appropriate to
the pCf. This case is particularly interesting since, in contrast with
other wall-bounded flows such as Poiseuille flow or the laminar
boundary layer (Blasius) flow, it completely lacks linear instability
and the absence of downstream advection makes it practical to observe
the long term dynamics of spots, a crucial element of the sub-critical
transition to turbulence. 

Contrasting with the regime of developed
turbulence taking place at large $R$, which requires a refined
space-time resolution able to account for a cascade toward small
scales, the transitional regime around $R_{\rm g}$ involves structures
that appear to be coherent and large, i.e. occupy the full
gap between the plates \cite{Betal98a}.  In turn, a model in terms of
a few well-chosen modes with low wall-normal resolution should be
sufficient for describing spots and understanding their dynamics. The
amplitudes of these modes would then be taken as functions of the
in-plane coordinates. This is the reason why we speak of {\it
  2.5-dimensional models}. In this terminology, the number~$2$ stands
for the full in-plane space dependence ($x,z$) and the suffix~.5
suggestively expresses that the dependence on the third, cross-stream,
coordinate $y$ is only partly taken into account through low order
truncation.  This strategy has been developed previously for
easy-to-treat but unphysical {\it stress-free\/} boundary conditions at
the plates bounding the flow \cite{ML00} yielding interesting results
but at unrealistically low Reynolds numbers \cite{MD01,Ma04}. Here we
focus on realistic {\it no-slip\/} boundary conditions. In the next
section, we recall the primitive equations and give a hint on how the
Galerkin method is developed. Next we turn to the explicit derivation
of the model (the no-slip and stress-free models are compared at a
formal level in \S\ref{S-COMP}). The numerical
implementation is described in \S\ref{S-NUM}. Our main results are
presented in Section~\ref{S-RES}, essentially bearing on the globally
sub-critical character of the transition to turbulence (\S\ref{S-SUB}).
The behavior of associated transients is then studied by
experiments where the system is quenched from a turbulent state at
large $R$ towards ever decreasing values of $R$ (\S\ref{S-GST}). These
results are discussed and some conclusions are drawn in Section~\ref{S-CONC}.

\section{The 2.5D models of 3D transitional flow}
\label{25mod}
\subsection{Primitive perturbation equations}

The Navier-Stokes equation and continuity condition for an incompressible
flow read:
\BA
\label{E-NS1}
\D_t\V+\V\cdot\bnabla\V&=&-\bnabla p+
\nu\bnabla^2\V+{\bf f}\,,\\
\label{E-NS2}
\bnabla\cdot\V&=&0\,,
\EA
where $\V\equiv(u,v,w)$, $p$ is the pressure 
and $\nu$ is the kinematic viscosity. Further, $\bnabla^2$ denotes the
three-dimensional Laplacian and the term ${\bf f}$ accounts for a driving
bulk force, if necessary.

Two types of boundary conditions can be considered, either the easy-to-handle
but unrealistic stress-free (sf) conditions, or the natural and
realistic no-slip (ns)
conditions. In the sf case, the fluid is supposed to slip on the walls
so that the flow has to be maintained by a fictitious bulk force
$\mathbf f \propto\sin(\beta y/h)$ with $\beta=\pi/2$, 
further adjusted to produce the basic velocity profile $U_{\rm
  b}^{\rm sf}(y)=U_{\rm p}\sin(\beta y/h)$.
In contrast, with ns boundary conditions, the motion of the plates is
able to drive the flow, which (when laminar) results in the linear
velocity profile $U_{\rm b}^{\rm ns}(y)=U_{\rm p} y/h$, where
$2U_{\rm p}$ is the relative speed of the plates, and $2h$ is the gap
between them.

In the following we use dimensionless quantities. According to common
usage, lengths and speeds are scaled with $h$, and $U_{\rm p}$,
respectively. In the ns case, the basic velocity profile thus reads
$U_{\rm b}^{\rm ns}=y$ for $y\in[-1,1]$, and the Reynolds number is
$R=U_{\rm p}h/\nu$.
The models are further written for the perturbation $(u',v',w',p')$
to the laminar basic flow that defines the ${\bf \hat x}$ direction,
${\mathbf v}_{\rm b} = U_{\rm b}\,{\bf \hat x}$, with
$U_{\rm b}=U_{\rm b}^{\rm sf}$ or $U_{\rm b}^{\rm ns}$, {\it i.e.}
$u=U_{\rm b}(y)+u'$, $v=v'$, $w=w'$, $p=p'$,
so that (\ref{E-NS1},\ref{E-NS2}) then read:
\BA
\NN
&&\D_t u'+u'\D_x u'+v'\D_y u'+w'\D_z u'\\
\label{E-NSpu}
&&\quad =-\D_x p'-U_{\rm b} \D_x u'-v' \mbox{$\frac{\rm d}{{\rm d}y}$} U_{\rm b}+R^{-1}\bnabla^2 u'\,,\\ 
\NN
&&\D_t v'+u'\D_x v'+v'\D_y v'+w'\D_z v'\\
\label{E-NSpv}
&&\quad =-\D_y p'-U_{\rm b} \D_x v'+R^{-1}\bnabla^2 v'\,,\\
\NN
&&\D_t w'+u'\D_x w'+v'\D_y w'+w'\D_z w'\\
\label{E-NSpw} 
&&\quad =-\D_z p'-U_{\rm b} \D_x w'+R^{-1}\bnabla^2 w'\,,\\ 
\label{E-NSpe} 
&&\,0=\D_x u'+ \D_y v'+\D_z w'\,.
\EA

The Galerkin method is a special case of a {\it weighted residual method\/}
\cite{Fin72}. It consists here in forcing the separation of in-plane and
wall-normal coordinates by expanding the perturbations ($u',v',w',p'$) onto
a complete basis of $y$-dependent orthogonal functions
satisfying the BCs with
{\it amplitudes\/} dependent on $(x,z,t)$. The equations of motion are then
projected onto the {\it same\/} functional basis, using the canonical
scalar product.
The main modeling step is then performed when truncating these expansions at a
low order and keeping the corresponding number of residuals in order to get
a  consistent and closed system governing the amplitudes retained.

\begin{figure*}
\begin{center}
\includegraphics[width=0.31\textwidth,clip]{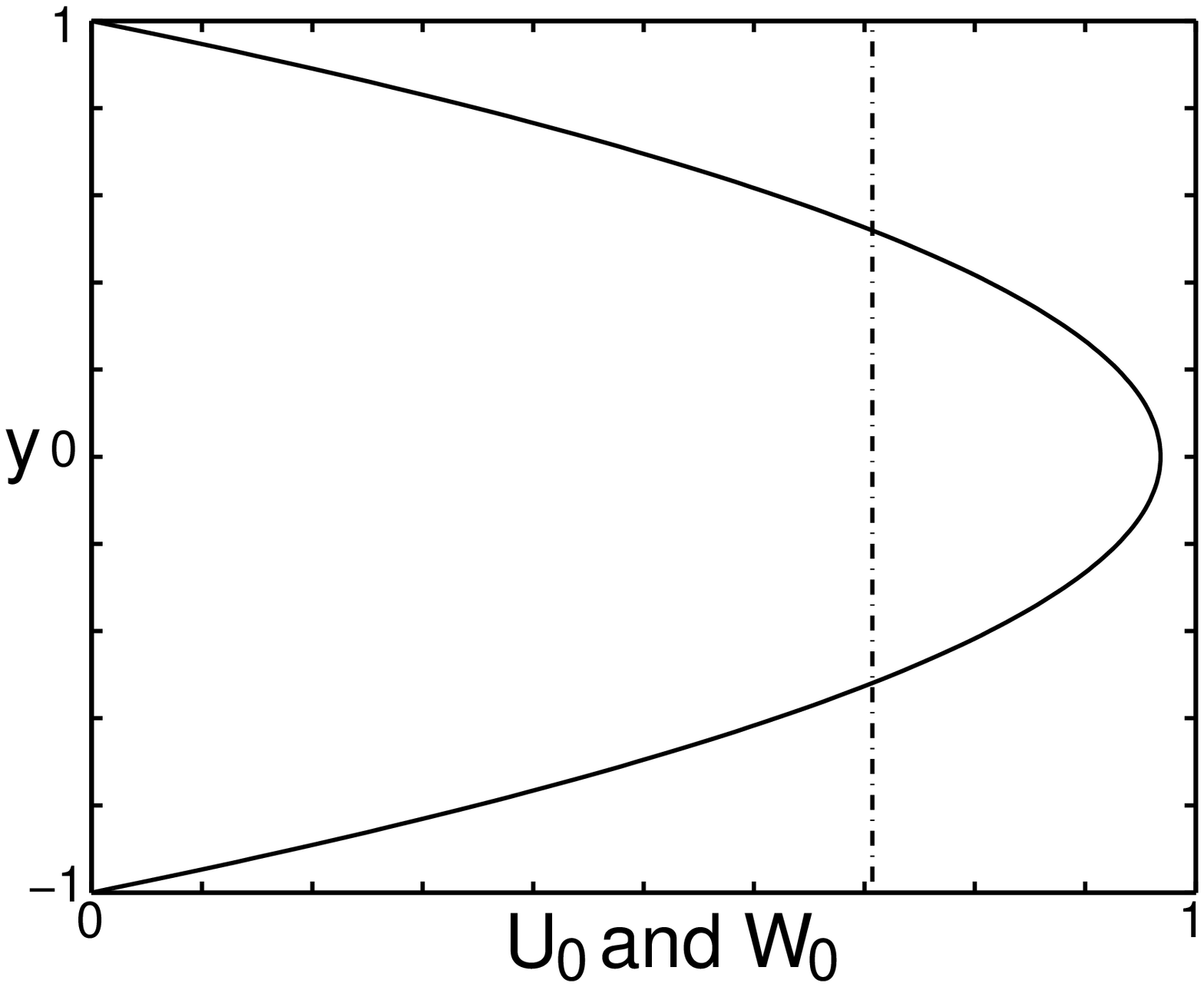}\hfill
\includegraphics[width=0.31\textwidth,clip]{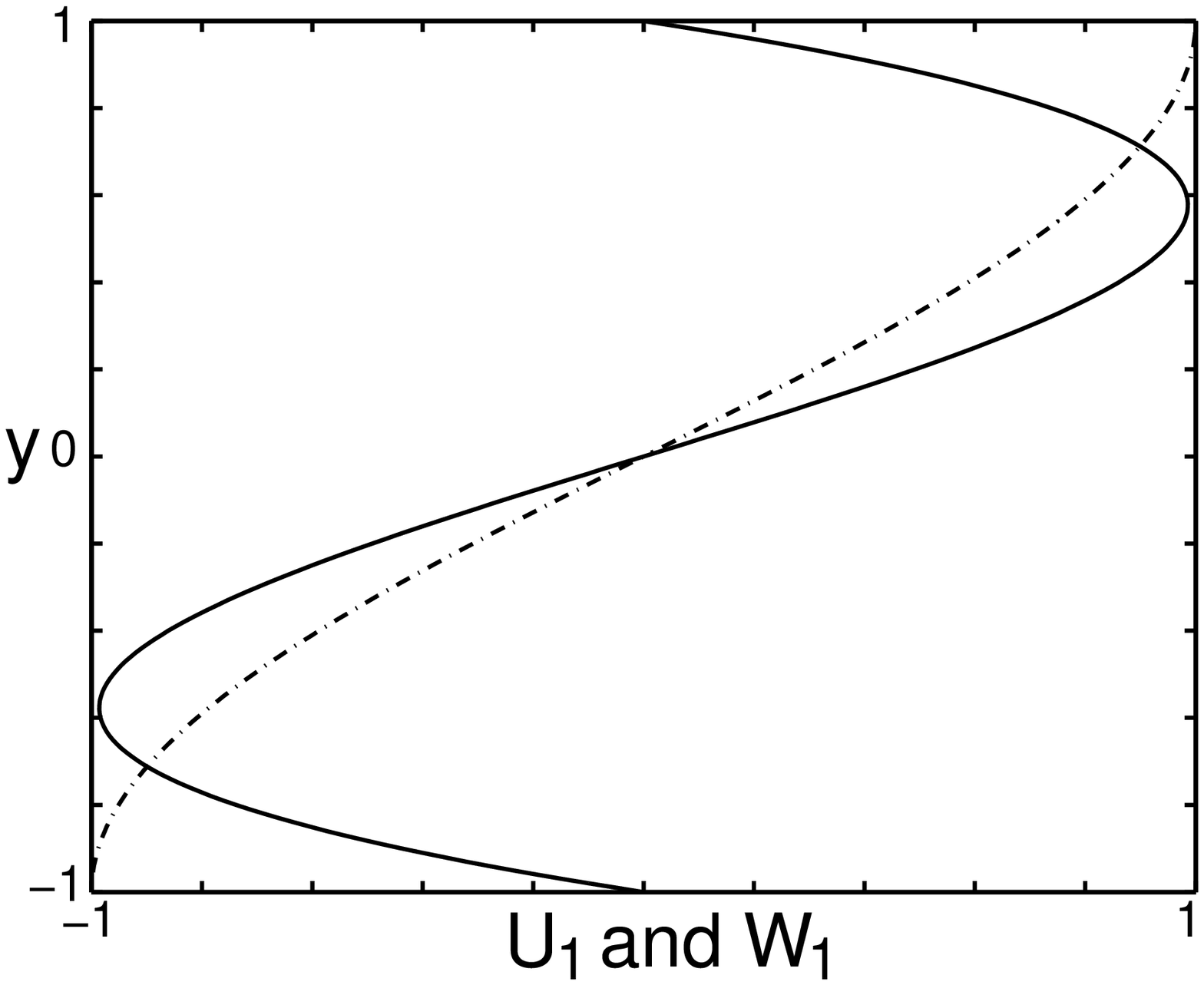}\hfill
\includegraphics[width=0.31\textwidth,clip]{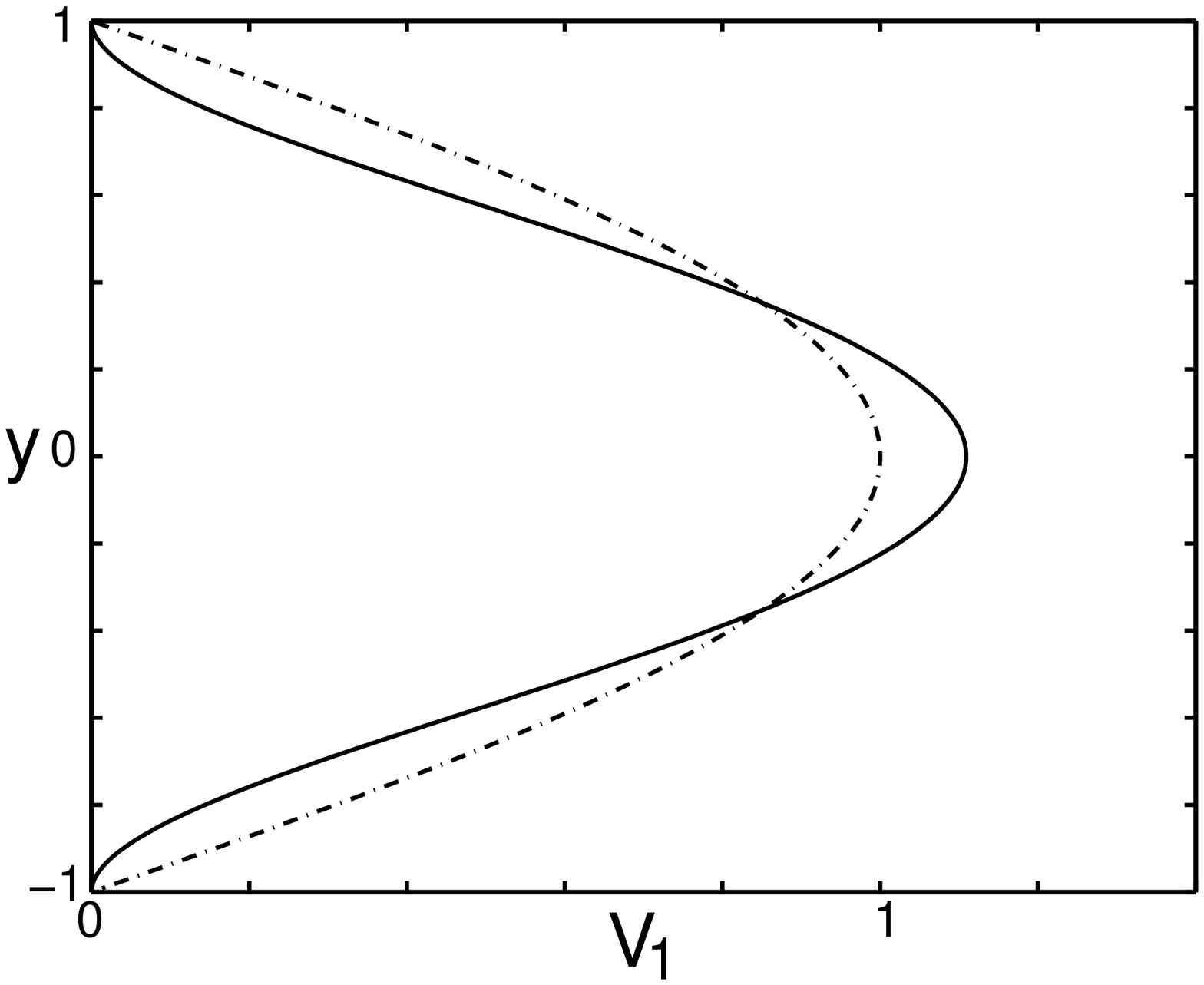}
\end{center}
\caption{Profiles of the basis functions used for the perturbation to
  the base flow in the derivation of the ns model (solid lines),
  compared to their counterparts for the sf model (dashed-dotted
  lines) using the same mean-square normalization: drift flow
  $B(1-y^2)$ vs. $1/\sqrt2$, first odd component $C y(1-y^2)$,
  vs. $\sin(\pi y/2)$, and first even component $A(1-y^2)^2$
  vs. $\cos(\pi y/2)$.}
\label{figp}
\end{figure*}

\subsection{The no-slip model at lowest order\label{S-nsl}}

The free-slip model introduced in \cite{ML00} and used more
extensively in \cite{Ma04} was developed by taking advantage of the
fact that trigonometric functions form a closed set under
differentiation and multiplication. 
Perturbations were taken of the form: $v'=V_1 \cos(\beta y)$ and%
\footnote{\label{fn1}Here the amplitude of the $y$-independent
  function is normalized differently than in~\cite{ML00,Ma04} in order
  to ease forthcoming comparisons.}
$\{u',w'\}=\sqrt{1/2}\{U_0,W_0\}+\{U_1,W_1\}\sin(\beta y)$ with
$\beta=\pi/2$ and the derivation of the model was straightforward.
Numerical simulations have shown that it displays the expected globally
sub-critical transition to turbulence but at unrealistically low Reynolds
numbers \cite{MD01}. This can be attributed to the underestimation of
viscous dissipation effects implied by the choice of sf boundary
conditions, in the same way as for convection where the sf threshold
is about 2/5 that for ns conditions ($27\pi^4/4$ vs. 1708).
The derivation of a model for the realistic case by
a similar Galerkin approach is harder because simple
trigonometric expansions do not satisfy the simultaneous
conditions: 
\BE
\label{E-Vnsbc}
v'(y=\pm 1)=\D_y v'(y=\pm 1)=0
\EE
obtained by combining the continuity equation (\ref{E-NSpe}) with the
conditions 
\BE
\label{E-UWnsbc}
u'(y=\pm1)=w'(y=\pm1)=0\,.
\EE
To overcome this difficulty we take a basis consisting of polynomials
in the wall-normal coordinate $y$, which also forms a family of
functions closed under multiplication and differentiation. A strict
Galerkin approach 
is followed \cite[Chap.~11, \S10]{GP71}, with projections defined using
the canonical scalar product:
$$
\langle f,g\rangle=\int_{-1}^{+1} f(y)g(y)\,{\rm d}y\,.
$$

From conditions (\ref{E-Vnsbc}) it is clear that the
expansion for $v'$ must be taken of the form \cite[p.~110]{Ma90}:
\BE
\label{E-Rm}
v'=(1-y^2)^2\sum_n V_{n+1} R_n(y)
\EE
where the $R_n(y)$, $n=0,1,...$ are polynomials of increasing degree $n$.
The basis for $u'$ and $w'$ is then obtained by requiring that
(\ref{E-NSpe}) 
be verified for each $n$ and that the basis be orthonormal. It can be
checked that these two conditions can be satisfied simultaneously and
that $u'$ and $w'$ expands as
\BE
\label{E-Qm}
\left\{u',w'\right\}=(1-y^2)\sum_n
\left\{U_n,W_n\right\} S_n(y)
\EE
where the expression of $S_n(y)$ derives from that of $R_n(y)$ using
the rules stated above. The Galerkin procedure isolates the dependence
of the velocity field on the cross-stream coordinate 
$y$ and, in expansions (\ref{E-Rm},\ref{E-Qm}), $U_m$, $V_m$, $W_m$
are just functions of the in-plane coordinates $x$, $z$, and time $t$.

It is quite simple to restrict consistently
to the lowest possible order, {\it i.e.} keeping only functions associated
to $U_0$, $W_0$, $U_1$, $V_1$, and $W_1$ as for the lowest order stress-free
model. This is due to the parity properties of the functions involved that
automatically guarantee the orthogonality of the different contributions
to the velocity field. Our expansion will thus be based on
\BE
\label{E-Vbc1}
v'=V_1 A(1-y^2)^2\,,
\EE
as the equivalent of $V_1\cos(\beta y)$ and accordingly:
\BE
\label{E-UWbc1}
\{u',w'\}= \{U_0,W_0\}B(1-y^2) + \{U_1,W_1\} Cy(1-y^2)\,,
\EE
in replacement of $\frac{1}{\sqrt2} \{U_0,W_0\} + \{U_1,W_1\}\sin(\beta y)$.
We need not worry about expanding the pressure since, at each stage of
the Galerkin projection, pressure contributions ---here $P_0$ and
$P_1$--- introduce themselves as Lagrange multipliers serving to fulfill
the successive projections of the continuity equation (\ref{E-NSpe}),
here just the projections on the lowest order even and odd velocity basis
functions (\ref{E-UWbc1}). Physically, the contribution  to the flow
identified by $(U_0,W_0)$ corresponds to the streak component induced by
the lift-up mechanism acting through the component $V_1$ of the contribution
$(U_1,V_1,W_1)$, a part of which corresponds to streamwise vortices,
flow structures that play an important role in the self-sustainent of
wall-turbulence. Going beyond lowest consistent truncation
order is tractable but somewhat tedious and we believe that the lowest
order model contains all the large scale features present in the
experiment at a qualitative level and can account for them at a
semi-quantitative level.

Throughout the derivation, integrals to be computed are of the form:
$$
J_{n,m}=\int_{0}^{1} y^{n}(1-y^2)^{m} {\rm d}y
=\sum_{k=0}^{m} {k\choose m}\frac{(-1)^k}{2k+n+1}\,,
$$
where the $k\choose m$ are the binomial coefficients.
For example, the normalization constants are given by:
$A^2=1/2J_{0,4}=315/256$, $B^2=1/2J_{0,2}=15/16$, and
$C^2=1/2J_{2,2}=105/16$.

Amplitudes introduced in (\ref{E-Vbc1},\ref{E-UWbc1}) are all functions
of $x$, $z$, and $t$. 
The drift-flow component ($U_0,W_0$) now has a plane Poiseuille
profile, in close correspondence  with what happens in the
Rayleigh-B\'enard case, as first noticed by Siggia and Zippelius \cite{SZ81}
and subsequently exploited to derive the generalized Swift--Hohenberg
model described in~\cite{Ma83}. Figure~\ref{figp} compares the profiles
of the basis functions for the free-slip and no-slip cases used at lowest
consistency order.

According to the prescriptions of the Galerkin method, we insert
the assumed expansions (\ref{E-Vbc1},\ref{E-UWbc1}) in the continuity equation
(\ref{E-NSpe}), then we multiply it by $B(1-y^2)$ and integrate over
the gap, which extracts its even part:
\BE
\label{E-C0}
\D_x U_0+\D_z W_0=0\,.
\EE
In the same way, multiplying (\ref{E-NSpe}) by
$Cy(1-y^2)$ and integrating extracts the odd part that reads:
\BE
\label{E-C1}
\D_x U_1+\D_z W_1=\beta\, V_1\,,
\EE
where the quantity $\beta=\sqrt3=\beta^{\rm ns}$ has the same order of magnitude as
the one appearing in the stress-free model
$\beta\,^{\rm sf}=\pi/2$ (see \S\ref{S-COMP}).

Doing appropriate manipulations on (\ref{E-NSpu}) and (\ref{E-NSpw}),
we obtain the even parts as
\BA
\NN
\D_t U_0 +N_{U_0}&=&-\D_x P_0 -a_1\D_x U_1 -a_2 V_1\\
\label{E-U0}
&&\quad\mbox{} +R^{-1}  \left(\Delta -\gamma_0\right)U_0\,,\\
\NN
\D_t W_0 +N_{W_0}&=&-\D_z P_0 -a_1\D_x W_1\\
\label{E-W0}
&&\quad\mbox{} +R^{-1} (\Delta -\gamma_0)W_0\,,
\EA
with $a_1=1/\sqrt7$, $a_2=\sqrt{27/28}$, $\gamma_0=5/2$, and
\BA
\NN
N_{U_0}&=&\alpha_1 (U_0 \D_x U_0+W_0 \D_z U_0)\\
\label{E-NU0}
&&\mbox{}+\alpha_2(U_1\D_xU_1+W_1\D_zU_1+\beta'V_1U_1)\,,\\
\NN
N_{W_0}&=&\alpha_1 (U_0\D_x W_0+W_0 \D_z W_0)\\
&&\mbox{}+\alpha_2(U_1\D_xW_1+W_1\D_zW_1+\beta'V_1W_1)\,,\label{E-NW0}
\EA
with $\alpha_1=3\sqrt{15}/14$, $\alpha_2=\sqrt{15}/6$,
$\beta'=3\sqrt3/2=\frac32\beta$.
In (\ref{E-U0},\ref{E-W0}) and
below $\Delta$ denotes the two-dimensional Laplacian
$\D_{xx}+\D_{zz}$.

In the same way, the odd part of (\ref{E-NSpu}) and (\ref{E-NSpw}) read:
\BA
\NN
\D_t U_1 +N_{U_1}&=&\mbox{}- \D_x P_1-a_1\D_x U_0\\
\label{E-U1}
&&\quad\mbox{} +R^{-1} (\Delta-\gamma_1)U_1\,,\\
\NN
\D_t W_1 +N_{W_1}&=&\mbox{}-\D_z P_1 -a_1\D_x W_0\\
\label{E-W1}
&&\quad\mbox{}+R^{-1} \left(\Delta-\gamma_1\right)W_1\,,
\EA
with $\gamma_1=21/2$ and  
\BA
\NN
N_{U_1}&=&\alpha_{2} (U_0\D_x U_1+U_1 \D_x U_0\\
\label{E-NU1} &&\mbox{}+W_0 \D_z U_1+W_1 \D_z U_0-\beta''V_1 U_0)\,,\\
\NN
N_{W_1}&=&\alpha_{2} (U_0 \D_x W_1+U_1 \D_x W_0\\
\label{E-NW1}&&\mbox{} +W_0\D_z W_1 +W_1 \D_z W_0-\beta''V_1 W_0)\,,
\EA
with $\beta''=\sqrt3/2=\frac12\beta$.

Finally, the equation for $V_1$ is obtained in the same way by
projecting (\ref{E-NSpv}) onto (\ref{E-Vbc1}), which yields:
\BA
\label{E-V1}
\D_t V_1 +N_{V_1}&=&-\beta P_1 +R^{-1} (\Delta -\gamma_1')V_1,\\
\label{E-NV1}
N_{V_1}&=&\alpha_3(U_0\D_xV_1 +W_0\D_zV_1)\,,
\EA
with $\gamma_1'=\beta^2$ and $\alpha_3=5\sqrt{15}/22$.

\begin{table*}
\caption{Comparison of coefficients for the sf and ns models at lowest
  order. Definitions are those introduced for the no-slip model and
  further identified term-by-term in the stress-free model (the
  factors $\sqrt2$ all stem from the normalization convention for the
  $y$-independent mode associated to amplitudes $\{U_0,W_0\}$, see
  Figure~\ref{figp} and the text relative to the sf basis).}
\begin{center}
\begin{tabular}{|c|c|c|c|c|c|c|c|c|c|c|c|c|c|}
\hline
coefficient & $\beta$ & $\beta'$ & $\beta''$
& $a_1$ & $a_2$\\
\hline
stress-free (sf) & $\pi/2$ & $\beta\,^{\rm sf}$ & 0  & $1/\sqrt2$&
${a_1}^{\rm sf}\beta\,^{\rm sf}$\\
\hline
no-slip (ns)& $\sqrt3$ 
            & $\sfrac32\beta\,^{\rm ns}$ & $\sfrac12\beta\,^{\rm ns}$ 
& $1/\sqrt7$ & $\sfrac32  {a_1}^{\rm ns} \beta\,^{\rm ns}$\\
\hline
sf/ns  & 0.907 & 0.605 & 0 & 1.871 & 1.131\\
\hline
\end{tabular}
\vspace{1ex}

\begin{tabular}{|c|c|c|c|c|c|c|c|c|c|c|c|c|c|}
\hline
coefficient & $\alpha_1 $ & $\alpha_2$ & $\alpha_3$ & $\gamma_0$ &
$\gamma_1$ & $\gamma_1'$ \\
\hline
stress-free (sf) & $1/\sqrt2$ & $1/\sqrt2$
            & $1/\sqrt2$ & 0 & $(\beta^{\rm sf})^2$ & $(\beta^{\rm sf})^2$\\
\hline
no-slip (ns)& $3\sqrt{15}/14$ & $\sqrt{15}/6$ & $5\sqrt{15}/22$ & 5/2  & 21/2
            & $(\beta^{\rm ns})^2$ \\
\hline
sf/ns       & 0.852 & 1.095 & 0.803 & 0 & 0.235 & 0.822 \\
\hline
\end{tabular}
\end{center}
\label{tabc}
\end{table*}
Notice that coefficients $\alpha_j$ behave as constants in the model,
while the $\beta$s are analogous to wave-vectors accounting
for cross-stream differentiation applied to the corresponding terms.
In the same spirit one could remark that, in (\ref{E-U0}),
$a_2=\beta' a_1=\frac32\beta a_1$. 
Another point worth mentioning is that, by construction, the
nonlinearities in the model preserves the energy conservation
properties of the advection term $\V\cdot\bnabla\V$.
This fact is indeed important because the nonlinear evolution is
formally governed by a quadratic expression which may produce
singularities in a finite time. Energy conservation can be verified by
computing the evolution of $E_{\rm t}=E_0+E_1=\frac12(U_0^2+W_0^2) +
\frac12(U_1^2+V_1^2+W_1^2)$. This tedious
task can be by-passed by taking advantage of the identity:
$\V\cdot\bnabla\V\equiv\bnabla\left(\sfrac12\V^2\right)+\bomega\cdot\V$,
with $\bomega=(\bnabla\V)^{\rm t}-\bnabla\V$, where
$(\bnabla\V)^{\rm t}$ is the transpose of tensor $\bnabla\V$, which
leads to the rewriting of the advection terms given in the appendix.

\subsection{Comparison of stress-free and no-slip models\label{S-COMP}}

The no-slip model derived above appears
to be slightly more general than the stress-free model initially presented
in \cite{ML00}. The two models have exactly the same structure, apart
from the term multiplied by $\beta''$ which is absent in the
stress-free case. Furthermore, when normalized in
the same way, their coefficients are comparable, while the main
difference lies, as expected, in the terms associated to viscous
dissipation. Let us examine Table~\ref{tabc} in detail.

As already said, values found for the quantity $\beta$ in
(\ref{E-C1}), which measures the typical order of magnitude of
gradients in the wall-normal direction, are close to each other since
$\beta^{\rm ns}=\sqrt3\approx1.732$ and $\beta^{\rm sf}=\pi/2\approx1.571$.

Coefficients $a_{1,2}$ and $\alpha_{1-5}$ appearing in the other equations
are also not very different in the two models. From the sf/ns entries
in the table, one can see that ratios are all within a factor less
than two, with the exception of $\beta''$ which is zero in the
stress-free case owing to the special relations existing between
trigonometric 
lines. As far as non-normal effects and nonlinear interactions are
concerned, one may think that the two models are close to each other
and that the dynamics that they generate will be robust (the effects
of the differences could be studied in detail by considering
fictitious alternate models with arbitrary coefficients, i.e. not
derived from any systematic Galerkin approach). 

The most important difference between the two models shows up in the
expression of the viscous terms. Whereas in the stress-free model, the
drift velocity component $\{U_0,W_0\}$ can relax only through in-plane
modulations {\it via\/} the terms involving $\Delta$, an additional
damping is observed in the no-slip case with coefficient $5/2$ in
(\ref{E-U0},\ref{E-W0}) independent of the in-plane dependence
of that flow component. In the same way the damping of components
$\{U_1,W_1\}$ is much stronger in the no-slip case with coefficient
$\gamma_1=21/2$ in (\ref{E-U1},\ref{E-W1}), than in the stress-free case
with coefficient $(\pi/2)^2\approx2.41$. The ns wall-normal viscous
timescale is thus four times shorter than the sf one. Accordingly, if
turbulence can be sustained in the stress-free model at 
some value $R$, already independently of other sources of damping and
while the mechanisms are expected to retain most of their intensity as
discussed in the previous paragraph, a similar situation is expected
for about $4R$ in the no-slip model. Of course the argument is very
rough but it shows that one goes in the right direction since the
stress-free model is known to underestimate thresholds by a large
factor \cite{MD01}.

Another important feature of the no-slip model is that an
$(x,z)$-independent component of $U_1$ can be created as the flow
evolves, so that the model already contains a mean flow correction to
the base profile $U_b(y)=y$ even when truncated at lowest order.
In contrast, a similar correction formally appears only at a much
higher order in the stress-free model, as pointed out in~\cite{MFE04}.
In the latter case, velocity profiles attached to $U_1$ and $W_1$ are
indeed the same as that of the base flow and accordingly do not modify
its shape though an equivalent mean contribution also exists.
This feature is further examined in the next subsection.

\subsection{Mean flow correction}
Let us consider the average value
$$
{\mathcal D}^{-1}\int_{\mathcal D} U_1(x,z,t)\, {\rm d}x{\rm d}z
\equiv{\overline U}_1(t)
$$
of the streamwise perturbation velocity component $U_1$ over the domain
$\mathcal D$ at the boundary of which perturbations cancel, or else
periodic boundary conditions apply. The equation governing
${\overline U}_1(t)$ is easily obtained by averaging
(\ref{E-U1}) over the domain. On its right hand side, the only term that
remains in the average is the one that is not the gradient of some
quantity, i.e. $- \gamma_1R^{-1} {\overline U}_1$, accounting for viscous
dissipation. The evaluation of
$\int_{\mathcal D} N_{U_1}{\rm d}x{\rm d}z$
leads to ${\overline N}_{U_1}=
\alpha_2(\beta+\beta'') {\mathcal D}^{-1}\int_{\mathcal D}
U_0 V_1 {\rm d}x{\rm d}z$. This quantity ---the space-independent
contribution to the streamwise component of the Reynolds stress---  
is easily obtained after some simple algebra using the continuity
conditions (\ref{E-C0},\ref{E-C1}).
Gathering the results we get the evolution equation for the mean flow
correction as:
\BE
\label{E-U1bar}
\sfrac{\rm d}{{\rm d}t}{\overline U}_1
=\alpha_2(\beta+\beta''){\overline{U_0V_1}}
-\gamma_1R^{-1}{\overline U}_1\,.
\EE
As will be illustrated later (cf. Figure~\ref{fmf}), a fully turbulent
regime at steady state  will display a time-independent non-zero
---in fact negative--- correction, arising from the compensation of
the two terms on the right hand side.

The spanwise component of the correction ${\overline W}_1$ is governed
by the same equation as (\ref{E-U1bar}) provided that we make the
general change $U\mapsto W$. However ${\overline W}_1$
is expected to cancel on average for a turbulent regime at steady state
as a consequence of the $z\mapsto-z$ symmetry of the original
problem. The full set of perturbation equations
is indeed symmetric in the interchanges $x\leftrightarrow z$ and
$U\leftrightarrow W$ except for the presence of the
term $-a_2 V_1$  accounting for lift-up in (\ref{E-U0}), which has no
equivalent in its spanwise analogue (\ref{E-W0}). This is the only
term that singles out the streamwise direction, subsequently
introducing the difference between ${\overline U}_1$ and
${\overline W}_1$. A non-zero ${\overline W}_1$ could then only result
from an instability breaking that symmetry.%
\footnote{\label{fn2}Such a symmetry breaking was experimentally
  observed in the upper part of the transitional regime of pCf
  \cite{Petal02}, a range of Reynolds numbers which is out of reach
  of our present modeling due to its reduced cross-flow resolution.}

Whereas it is clear that no such correction exists for $V_1$ since
${\overline V}_1$ obviously cancels all the time (no net flux across
the layer for impermeable walls), we must also consider averaging the
fields $U_0$, $W_0$. The equation for ${\overline U}_0$ reads:
\BE
\label{E-U0bar}
\sfrac{\rm d}{{\rm d}t}{\overline U}_0
=\alpha_2(\beta-\beta'){\overline{U_1V_1}}
- \gamma_0 R^{-1}{\overline U}_0\,.
\EE
A similar equation can be obtained for ${\overline W}_0$ through
the change $U\mapsto W$. The source terms for
${\overline U}_0$ and ${\overline W}_0$ are however expected to be small
since, from the continuity condition (\ref{E-C1}), $U_1$ and $W_1$
are spatially out of phase with $V_1$. These terms can even be expected
to cancel statistically since otherwise they would generate net
contributions to the transfer of momentum in the $x$ and $z$ directions
and correlative finite dissipation terms $-\frac52 R^{-1} \overline
U_0$ and $-\frac52 R^{-1} \overline W_0$. As long as the global 
symmetries of the problem are not broken, net fluxes $\overline U_0$ 
and $\overline W_0$ are forbidden by the general $x\mapsto -x$
and $z\mapsto -z$ symmetries in the same way as
${\overline W_1}\ne0$ is forbidden by the $z\mapsto-z$ symmetry
(in contrast with $\overline U_1$ which has its own source term).

It can be noticed that, in the stress-free case, mode `0' has no
source term for accidental reasons linked to trigonometric relations
between base functions (i.e. $\beta=\beta'$). Furthermore, that mode
has no cross-stream dependence so that it 
cannot be ironed out by viscous dissipation. Accordingly,
${\overline U}_0$ and ${\overline W}_0$ retain their initial values
which, in turn, can freely be set to zero due to the in-plane
Galilean invariance specific to stress-free boundary conditions.

\section{Numerical implementation\label{S-NUM}}

The equations derived above have been implemented on domains of
various streamwise and spanwise sizes $L_x$ and $L_z$ with
periodic boundary conditions, which allowed us to develop a standard
Fourier pseudo-spectral numerical scheme \cite{GO77}. The diagonal
viscous terms were integrated in spectral space exactly. The nonlinear
advective terms and the non-normal terms arising from linearization
around the base flow were evaluated in physical space and integrated
in time using a second order Adams--Bashforth scheme also used to
treat the average velocity components $\overline U_j, \overline W_j$,
$j=0,1$. Fast Fourier transforms were used to pass from spectral to
physical space and {\it vice versa\/}. Convergence of the results was
checked by taking different time steps $\delta t$  and space steps
$\delta x,\delta z$. In the range of Reynolds numbers under
consideration, $\delta t=0.01$ and $\delta x=\delta z=0.25$ were
found appropriate and used throughout the study.

Simulations were performed in boxes with sizes ranging from
$\mathcal D=L_x\times L_z=8\times8$ up to $64 \times 64$, although the
model was originally designed for wide domains.  A few short control
runs were performed for $\mathcal D$ up to $128\times128$.
At any rate, results presented here are mostly
dedicated to checking that the no-slip model captures the essential
features of the globally sub-critical nature of the transition to
turbulence in pCf, as a prerequisite to future work. In fact,
the sizes considered here are already several times larger than the
expected streak spacing $\lambda_z\sim6$ 
 \cite{Betal98a}. In the terms of chaos
theory, we would speak of ``weak confinement'' (see \cite[Chap.~8]{Ma90}
and below \S\ref{S-EXT}) in contrast with ``strong confinement''
legitimating the reduction to low dimensional differential systems.

Obtained with limited computing power, simulation results
presented here have already their own interest as will be shown in
\S\ref{S-GST} but let us emphasize the fact that accepting low
cross-stream resolution was the price to be paid for enlarging the
domains up to sizes that are much larger than what 
can be dealt with in direct simulations with highly resolved
cross-stream dependence.%
\footnote{\label{fn3}In their study of the relaxation of oblique
  turbulent bands 
  observed in the experiments \cite{Petal02}, Barkley and Tuckerman
  \cite{BT05} performed direct simulations in a quasi-one-dimensional
  box making an angle with the streamwise direction, very long in the
  direction perpendicular to the bands but a few MFUs long in the
  direction parallel to them, which represents an intermediate stage.} 

Two further points are worth mentioning about the implementation
of the model. First, the continuity conditions (\ref{E-C0},\ref{E-C1})
are dealt with by introducing appropriate stream functions $\Psi_0$,
$\Psi_1$ and velocity potentials $\Phi_1$ such that:
\BE
\label{E-Psi0}
U_0={\overline U}_0-\D_z \Psi_0\,, \qquad W_0={\overline W}_0+\D_x \Psi_0\,,
\EE
and
\BE
\label{E-PsikPhik}
U_1={\overline U}_1+\D_x \Phi_1-\D_z \Psi_1\,, \quad
W_1={\overline W}_1+\D_z \Phi_1+\D_x \Psi_1\,,
\EE
with
\BE
\label{E-Phik}
\beta V_1=\Delta \Phi_1\,.
\EE
Though not necessary on general grounds, the introduction of the
additional quantities ${\overline U}_0$, ${\overline W}_0$,
${\overline U}_1$, and ${\overline W}_1$, is forced by our choice of
in-plane periodic boundary conditions for the fields $\Psi_0$, $\Psi_1$,
and $\Phi_1$. This is because these uniform components are
generated by contributions to the potential $\Phi_1$ and stream-functions
$\Psi_0$, $\Psi_1$ that would vary linearly with $x$ and $z$, a spatial
dependence that is precluded by their representations as
Fourier series expansions inherent in our
computational approach.

The velocity contributions derived from these equations are next
reintroduced in the expression of the nonlinearities when needed.
The equations governing $\Psi_0$ and $\Psi_1$ are obtained
by cross-differentiating and subtracting the equations for the
velocity components in the usual way. Taking the divergence
of these equations yields an equation for the pressure which is next
used to determine the potential part of the velocity field accounted
for by $\Phi_1$. The main advantage of introducing the fields
$\Psi_0$, $\Psi_1$, and $\Phi_1$ is that one can construct arbitrary
but physically relevant divergence free velocity fields as initial
conditions since this characteristic is built in their definitions.
The uniform corrections $(\overline U_1,\overline U_0,
\overline W_1,\overline W_0)$
are computed in parallel by integrating (\ref{E-U1bar},\ref{E-U0bar})
and their $W$ counterparts.
The equations that have been numerically implemented are gathered
in the appendix.

\section{Results\label{S-RES}}
\subsection{Global sub-criticality\label{S-SUB}}

\begin{figure}
\begin{center}
\includegraphics[width=0.66\textwidth,clip]{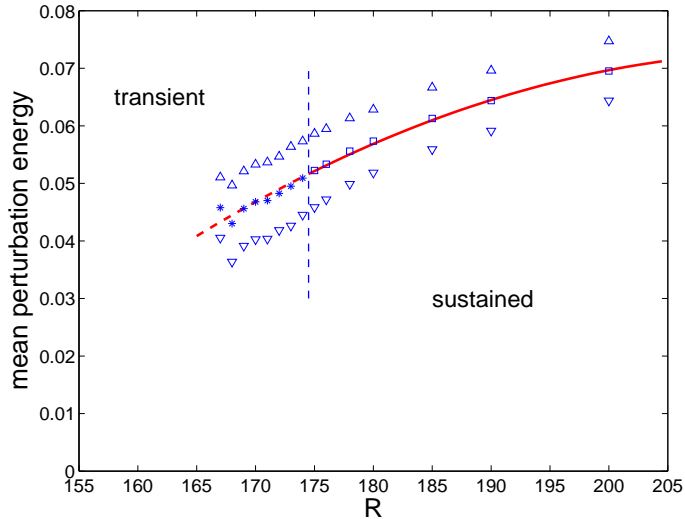}
\vspace*{-4ex}
\end{center}
\caption{Variation of the mean perturbation energy {\it per\/} unit
  surface with $R$. Evidence of a global stability threshold $R_{\rm
    g}\approx 175$. Values corresponding to squares $R>R_{\rm g}$
  correspond to sustained regimes, those corresponding to transients
  are marked with asterisks. Up and down triangles indicate the
  root-mean-square amplitudes of instantaneous fluctuations around
  the corresponding means. The computational domain is $\mathcal
  D\equiv32\times\nobreak32$.\label{ftt}}
\end{figure}

Global sub-criticality of the model is depicted in Figure~\ref{ftt}
which displays the behavior of the mean perturbation energy
{\it per\/} unit surface as a function of the Reynolds number $R$. A
discontinuous transition can indeed be observed around
$R_{\rm g}\approx175$. Whether $R_{\rm g}$ represents a real
threshold (called {\it global stability\/} threshold in previous
studies, hence the subscript `g') or just marks some crossover is
presently a debated topic for plane Couette flow \cite{BC98,Hetal06} as
well as for Poiseuille flow in a circular pipe which is also a
linearly stable flow that becomes turbulent under finite amplitude
perturbations \cite{Hetal06,PM06,WK07}.
For $R>R_{\rm g}$ spontaneous collapse of the
uniformly turbulent state was not observed, even when pursuing the
simulation for very long durations. For example, at $R=175$ the
turbulent state persisted over more than $3\times10^5$ time units.%
\footnote{\label{fn4}For experiments in water with a gap $2h=7$mm, in the
  transitional range, $R\sim315$, the time unit (turnover time
  $h/U_{\rm p}$) is $3.8\times10^{-2}$s, so that the time length
  $\sim 3\times10^5$ would correspond to about 3 hours, much longer
  than what could be done in the laboratory under sufficiently well
  controlled conditions \cite{Betal98b}.} 
On the other hand, below $\approx174.5$,
the laminar state was systematically obtained but at the end of
turbulent transients that could be of variable lengths. This result was
obtained by combining {\it quench\/} experiments in which a state
prepared at $R=R_{\rm i}=200$ was allowed to evolve after $R$ had been
suddenly decreased to some lower value $R_{\rm f}$ and {\it adiabatic\/}
experiments where $R$ was varied by small increments and the initial
state at the new lower value of $R$ was taken as the final state of the
experiment at the previous value. The persistent
turbulent regime at $R=175$ was obtained in this second way, which
---according to us--- indicates a lower stability
bound for the turbulent state at least over the corresponding time
interval and for the considered size ($L_x=L_z=32$) since the
procedure minimizes the perturbation brought to the flow, especially
regarding its uniform component $\overline U_1$. In contrast, above
$R_{\rm g}$ the initial finite perturbation brought to the system in
quench experiments may be sufficient to make it leave the attraction
basin of the turbulent state and return to the laminar state. Indeed,
if the turbulent regime is a genuine attractor, its attraction basin
shrinks to zero as $R$ approaches $R_{\rm g}$ from above and, with it, the
size of the perturbation brought to the turbulent state necessary to make it
decay toward the laminar state, in particular that linked to the fact that
$\overline U_1$ is not at equilibrium for the new value of $R$
after the quench.
At any rate, it appears that the range of interest for the
transition is no longer as low as in the sf case but
approaches the experimental value \cite{Betal98b}.

\begin{figure}
\begin{center}
\includegraphics[width=0.66\textwidth,clip]{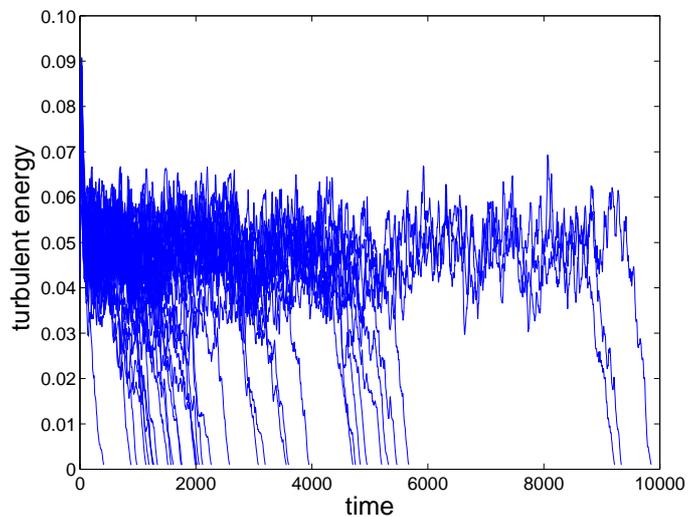}
\end{center}
\caption{Total perturbation energy {\it per\/} unit surface
$E_{\rm tot}$ as a function of time in a series of transients
from a turbulent flow prepared at $R_{\rm i}=200$ suddenly quenched at
$R_{\rm f}=171<R_{\rm g}\approx 175$.
\label{ftu}}
\end{figure}

A series of transients obtained in quench experiments from
$R_{\rm i}=200$ to $R_{\rm f}=171$ is illustrated in Figure~\ref{ftu}.
It can be seen that they end quite abruptly so that it makes
sense to define a conditional mean perturbation energy restricted to
the plateau value, before final decay.%
\footnote{\label{fn5}A similar observation was made by Faisst \& Eckhardt
  \cite{FE04} in numerical simulations of Poiseuille pipe flow. They
  pointed out that it was not possible to distinguish between
  sustained turbulence and turbulent transients before decay.}
Such conditional averages are
indicated as asterisks in Figure~\ref{ftt} while squares denote
averages corresponding to sustained turbulent regimes. The line
through the data points is a fit against a parabolic
expression that has no theoretical foundation but strongly suggests
that it goes smoothly through the critical value $R_{\rm g}$. In
particular, taking the mean perturbation energy as an order
parameter, we find no evidence of the proximity of a turning-point
bifurcation as was tentatively suggested in earlier
studies from analogies with elementary bifurcation theory \cite{Betal98b}.

The amplitudes of the fluctuations of the average perturbation energy
(conditional for transient states) are indicated by the up and down
triangles in Figure~\ref{ftt} and, again, no remarkable behavior
(i.e. divergence) can be noticed around $R\sim R_{\rm g}$. As
discussed in the next sub-section, their amplitude is comparatively
large only because, in this study, the size of the domain $\mathcal D$
has been chosen relatively small. The detailed statistics of the
transients' lifetimes, which is the main original result presented in
this paper, is studied in~\S\ref{S-GST}.

\subsection{Extensivity of the sustained turbulent regime\label{S-EXT}}

\begin{table*}
\caption{Perturbation energy {\it per\/} unit surface: time average and
  fluctuations
for
$R=200$.\label{text}}
\begin{center}

\begin{tabular}{|c|c|c|c|c|c|c|c|c|c|c|}
\hline
$\mathcal D: L_x\times L_z$ & $16\times16$ &
$32\times16$ & $32\times 32$ & $64\times 32$\\
\hline
$\langle E_{\rm tot}\rangle $& 0.06935 & 0.06932 & 0.06956 & 0.06980\\
\hline
$\sigma_{\rm Etot} (\times 10^3)$& 10.33 & 7.21 & 5.58 & 3.62\\
\hline
$\displaystyle\frac{\sqrt{L_xL_z}\,\sigma_{\rm Etot}^{L_xL_z}}{16\,\sigma_{\rm Etot}^{16\times16}}$& 1.0 & 0.987 & 1.079 & 0.992 \\[1ex]
\hline
\end{tabular}
\vspace{1ex}

\begin{tabular}{|c|c|c|c|c|c|c|c|c|c|c|}
\hline
$\mathcal D: L_x\times L_z$ & $64\times 64$ & $128\times64$ & $128\times128$\\
\hline
$\langle E_{\rm tot}\rangle $ & 0.06965 & 0.06959 & 0.06948\\
\hline
$\sigma_{\rm Etot} (\times 10^3)$ & 2.61 & 1.758 & 1.179\\
\hline
$\displaystyle\frac{\sqrt{L_xL_z}\,\sigma_{\rm Etot}^{L_xL_z}}{16\,\sigma_{\rm Etot}^{16\times16}}$ & 1.010 & 0.9629 & 0.9132\\[1ex]
\hline
\end{tabular}
\vspace{1ex}
\end{center}
\end{table*}

It has been argued elsewhere by one of us (PM in~\cite{MK05}) that the
modeling in terms of low-dimensional ordinary differential systems
\cite{Wa97} may give valuable hints only about the mechanisms of
turbulence sustenance well beyond the transitional regime but not
necessarily about the transitional regime itself, where
spatio-temporal behavior is involved. Indeed, the main underlying
hypothesis of this kind of modeling is that the space-time dependence of
the perturbations can be described by mixing a small number of modes
with frozen space dependence and time varying amplitudes. Specific
resonances between modes make 
the temporal properties of the system, notably the fractal 
properties associated to chaotic transients, excessively sensitive to
the physical size of the equivalent confined system. Furthermore, the
approach is intrinsically unable to deal with the growth or decay of
turbulence through the coexistence of laminar and turbulent domains
that fluctuate in space and time \cite{Po86}. When considering
this specific problem, it seems legitimate to require that, to be
appropriate, a model should display some kind of statistical
robustness when the size of the simulation domain is varied
\cite{Po85}. An indication that our model behaves appropriately in the
turbulent regime is obtained by considering the total perturbation
energy {\it per\/} unit surface and its fluctuations as the surface of
the simulation domain is increased. 
Table~\ref{text} displays our results.  A series of experiments
over domains $\mathcal D=L_x\times L_z$ was performed, with domain
surfaces multiplied by factors $n=1$, 2,\dots, 64, starting
from $\mathcal D=8\times8$ and for $R=200$.

For the smallest systems considered, turbulence was only transient
at $R=200$. Needing larger $R$ to get sustained turbulence is not
surprising since much dissipation is associated with the in-plane
space dependence forced by periodic boundary conditions at short
distances, which raises the thresholds for complex behavior. For
$\mathcal D=8\times8$, the domain can only fit a very small number of
structures so that the transient is better analysed in terms of
chaotic saddles in the context of {\it temporal chaos\/} \cite{Ec06}.
For $\mathcal D=16\times8$  turbulence is again transient but the
dynamics has a more spatio-temporal flavor than for $8\times8$,
hence closer to what is obtained at larger sizes.
Evidence that the turbulent regime at $R=200$ is {\it extensive\/} for
larger domains, in practice beyond $\mathcal D=16\times16$, is
obtained from the facts that 
({\it i\/}) the time average of $E_{\rm tot}$ is independent of the
size and ({\it ii\/}) the fluctuations around this average 
value decrease roughly as the inverse square root of the size.

The extensivity property is valid for the uniformly
turbulent regime 
but size effects may still be expected to affect the transition
process, especially in view of the presence of large scale flows
resulting from the space modulation of the mean flow correction to be
discussed in the next sub-section, as soon as strong inhomogeneities
are present. This study only suggests us a typical size below which
the temporal approach is better suited. 

\subsection{Mean flow in the sustained turbulent regime\label{S-MF}}

\begin{figure}
\begin{center}
\includegraphics[height=0.4\textwidth,clip]{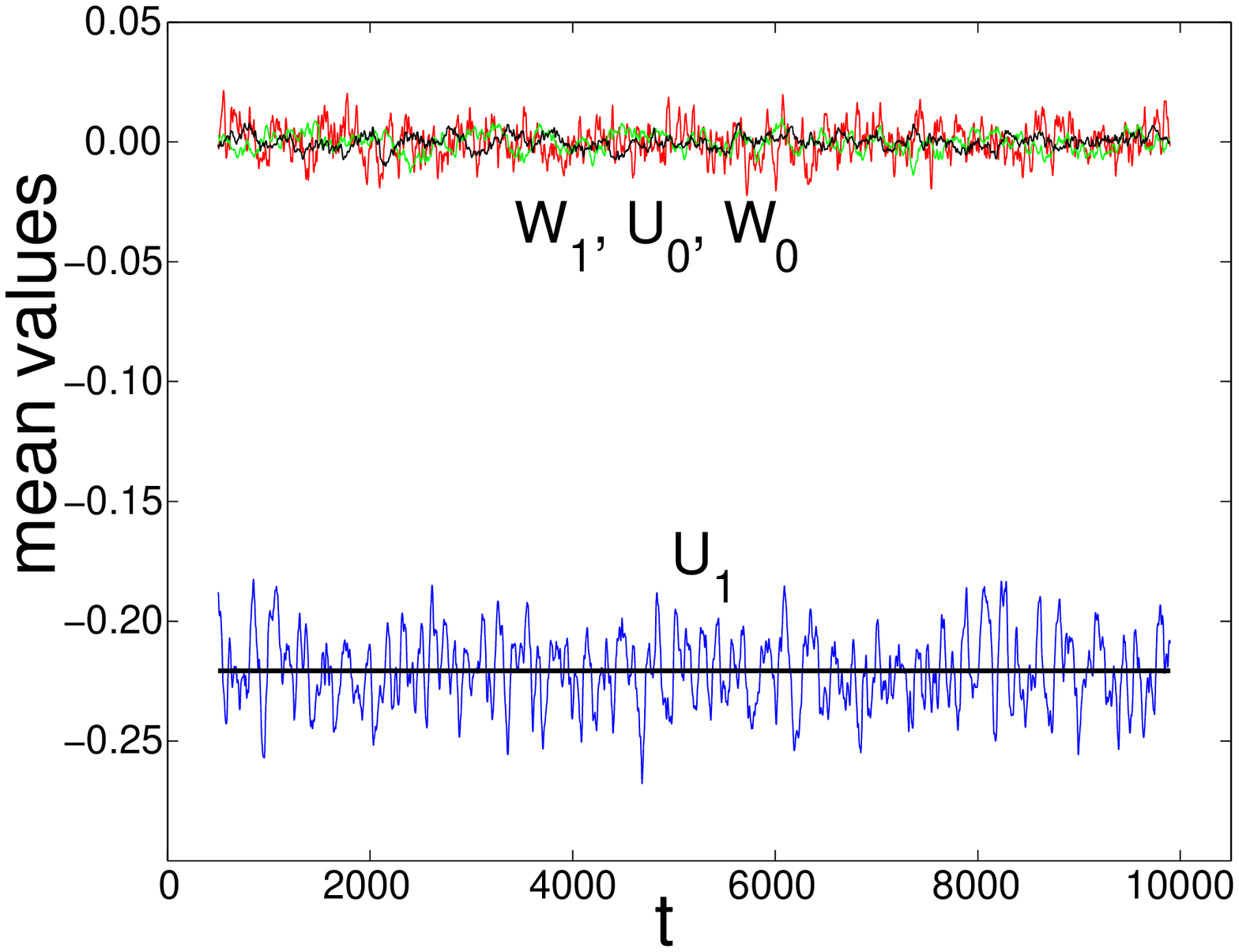}
\hfill
\includegraphics[height=0.4\textwidth,clip]{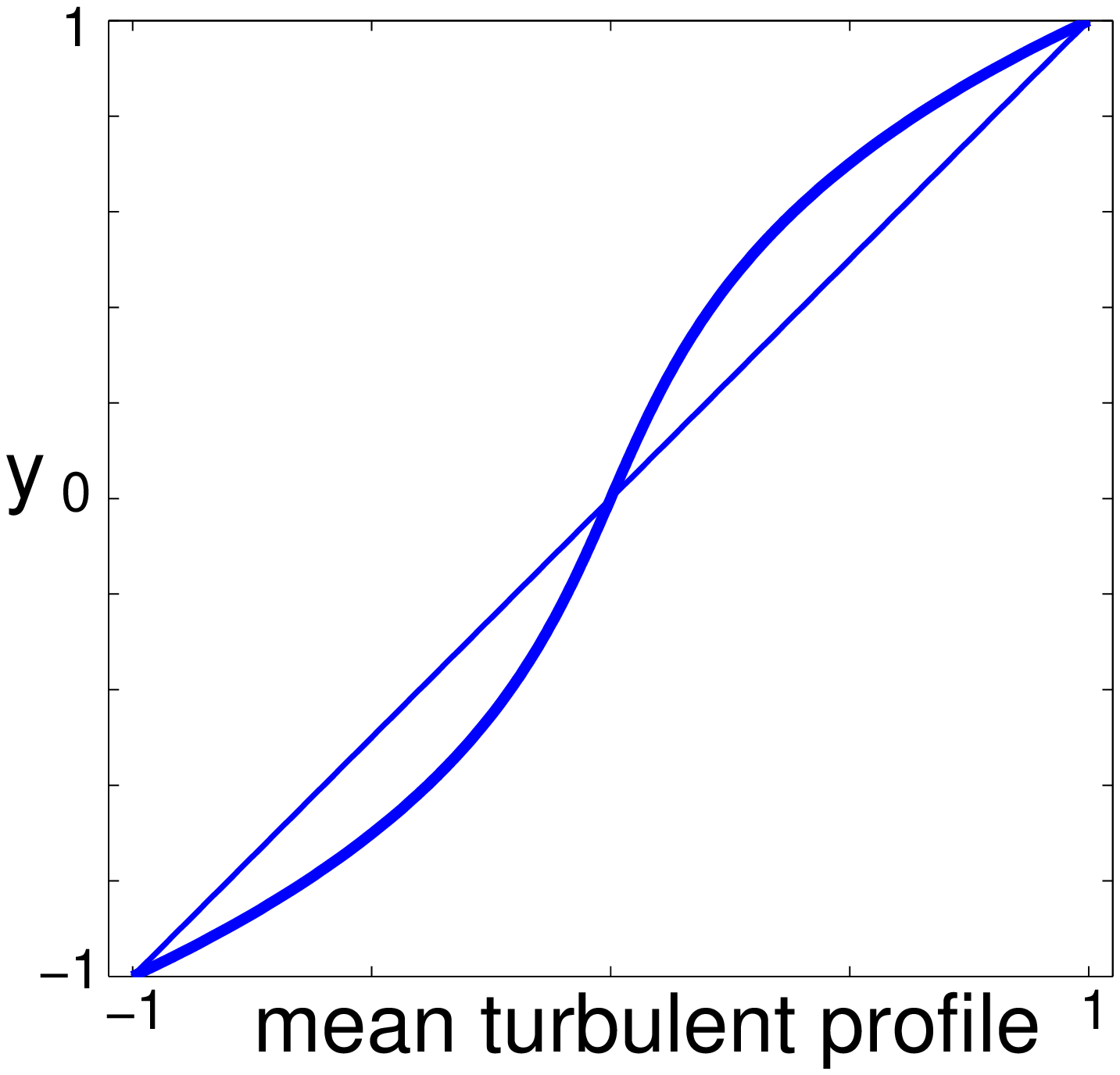}
\end{center}
\caption{Correction to the base flow for a sustained turbulent
regime obtained as described in the text ($R=200$, $\mathcal
D=32\times32$). Top panel: Traces of the average flow
components ${\overline U}_1$, ${\overline W}_1$, ${\overline U}_0$, and
${\overline W}_0$ as functions of time.
Bottom panel: Mean turbulent flow profile
$u_{\rm mf}=y+ {\overline U}_1Cy(1-y^2)$,
the value of ${\overline U}_1$ is obtained by averaging
the above relevant trace over $t\in[0.5,15]\times10^3$.
\label{fmf}}
\end{figure}

An important feature of our model is that it already contains corrections to
the base flow at the lowest truncation order.
Figure~\ref{fmf} illustrates the result of a specially
designed experiment at $R=200$ starting from an initial condition
taken from some fully turbulent state, in which all the components of
mode `1' were artificially turned off. During a brief transient that
has been cut off, the streamwise contribution to the mean flow
${\overline U}_1$ builds up. As seen in the figure, it is negative and
statistically constant, with time average
$\langle{\overline U}_1\rangle_t=-0.221$ and standard deviation
$\Sigma_{{\overline U}_1}=0.015$. In contrast, the
transverse contribution averages to zero:
$\langle\overline W_1\rangle_t=1.4\times10^{-4}$ with
$\Sigma_{{\overline W}_1}=6.9\times10^{-3} \>(\approx
50\,{\langle\overline W_1\rangle}_t$ for the specific numerical
simulation considered). At the same time,
$\overline U_0$ and $\overline W_0$ also average to zero:
$\langle{\overline U}_0\rangle_t=-2.4\times10^{-4}$, 
$\Sigma_{\overline U_0}=3.3\times10^{-3}$ and
$\overline W_0=-1.5\times10^{-5}$,
$\Sigma_{{\overline W}_0}=2.9\times10^{-3}$. The mean turbulent flow
profile is displayed in Figure~\ref{fmf} (bottom) as the superposition
of a correction $C y (1-y)$ with amplitude ${\overline U}_1$ to the
base flow $U_{\rm b}=y$, pointing out the expected formation of a
central region with reduced shear. It will be interesting to study
the patches of negative $U_1$ that appear during the growth/decay
of turbulent spots \cite{LM07}, as a local counter-part to the steady
state uniform correction in the sustained turbulent regime.

\subsection{Transients below the global stability threshold\label{S-GST}}

The global stability threshold $R_{\rm g}$ is best defined as the
value of the control parameter below which the nontrivial state, here
the turbulent regime, is unstable and thus cannot be sustained in the
long term. Accordingly the trivial state, here the laminar flow, is
the only possible equilibrium solution, whatever the initial
configuration (or accidental perturbations). Many studies have
been devoted to the transition from laminar
{\it to} turbulent flow, either
under the effect of natural fluctuations, upon triggering specific
localized perturbations, or upon modifying the base flow in a
well-controlled way (see the review in \cite{MD01}). In such
experiments, the turbulent state may be reached with finite
probability only due to the local stability of the base flow, which
explains the dispersion of results in early studies and some
sensitivity to the triggering process. In contrast, from the very
definition of $R_{\rm g}$, it may seem less ambiguous to start
{\it from} the turbulent side at large Reynolds
numbers and study its relaxation as $R$ is decreased. As already
mentioned one can consider either 
an {\it adiabatic\/} decrease of $R$ which minimizes the risk of
introducing dangerous perturbations, or else {\it quench\/}
experiments using the same methodology as the one which produced the
experimental value $R_{\rm g}\approx325$ \cite{Betal98a}.
The cumulative distribution of transient lifetimes was then
studied and found to display an exponential tail of the form
$\Pi(\tau'>\tau)\propto \exp(-\tau/\tau_R)$,
where $\Pi(\tau'>\tau)$ is the probability of observing a
transient with length $\tau'$ larger than some given time
$\tau$. In this expression $\tau_R$ is the characteristic decay time
of the distribution. If it was effectively exponential
from the start, then $\tau_R$ would be the mean transient length at the
corresponding value of $R$. From that estimate, it was proposed that
$\tau_R$ diverges to infinity as $R$ approached $R_{\rm g}$
from below as
\BE
\langle\tau_{\rm tr}\rangle\sim
(R_{\rm g}-R)^{-1}\,,
\label{E-CRIT}
\EE
see \cite{Betal98b,BC98}.
This behavior was recently questioned and a reexamination of the data
lead to the proposal of an indefinite exponential increase
\cite{Hetal06} though
the time scale computed in note~\ref{fn4} leaves little hope to
check the extrapolation at larger $R$ experimentally.
Our quench experiments with the ns model is a tentative contribution
to the debate.

\begin{figure}
\begin{center}
\includegraphics[width=0.66\textwidth,clip]{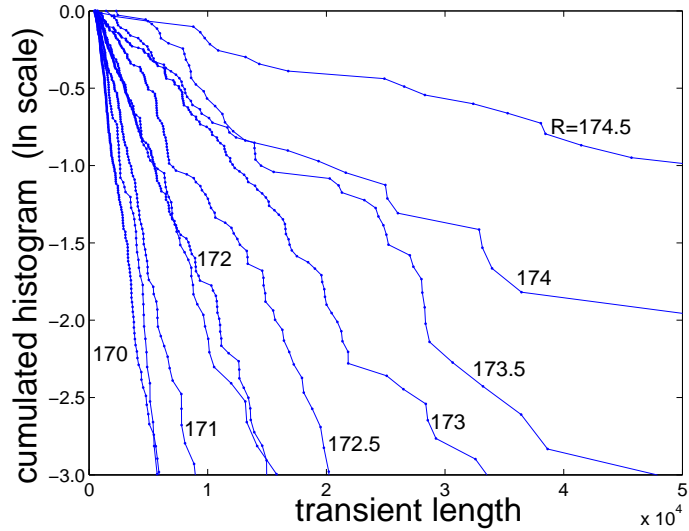}
\end{center}
\caption{Distribution of transient lifetimes in quench experiments
  with $R_{\rm i}=200$ and variable $R_{\rm f}$ in semi-log plot.\label{ftv}} 
\end{figure}

Experiments were performed by quenching the system from series of
uniformly turbulent states at $R=R_{\rm i}=200$. These states were obtained
from snapshots taken during a single long experiment and sufficiently
separated in time to be considered as independent turbulent initial
conditions. The Reynolds number was next suddenly decreased to a final
value $R=R_{\rm f}$. 
In order to be able to perform statistics over a large number of
independent trials, we took a domain size $\mathcal D$ that was
reduced as much as possible, though sufficiently large to preserve the
extensivity property of the uniformly turbulent regime. $\mathcal
D=32\times32$ was chosen but runs 
with $\mathcal D=128\times32$ or $128\times64$ were performed
for purposes of control or pattern imaging (see \S\ref{S-tr}).

The lifetimes of individual transients were determined as the times
needed to reach some cut-off value of the total perturbation energy
{\it per\/} unit surface $E_{\rm t}$ below which the system immediately and
irreversibly decays to the laminar state, as can easily seen
in Figure~\ref{ftu} displaying a typical set of experiments to be
analysed. The statistics were checked to be insensitive to the precise
value chosen for the cut-off provided that it was sufficiently below
$0.015$. Here the cut-off was uniformly taken equal to 0.0015 (see
also later, Fig.~\ref{endtr}). The very late stage of decay for
$E_{\rm tot}<0.0015$ appears to correspond to a fast exponential
viscous damping of streaks.

Below $R=175$, transients with variable lifetimes began to appear and
well-defined lifetime distributions were seen to develop, while
relaxation could be considered as ``immediate'' below
$R=R_{\rm u}\approx 164$ (notation introduced in \cite{Betal98b}).  
Upon approaching $R_{\rm g}\approx175$ from
below, using lin-log scales Figure~\ref{ftv} shows that the statistics of
transient lifetimes are indeed exponentially decreasing
and that the slope decreases as $R_{\rm f}$ increases.

\begin{figure}
\begin{center}
\includegraphics[width=0.49\textwidth,clip]{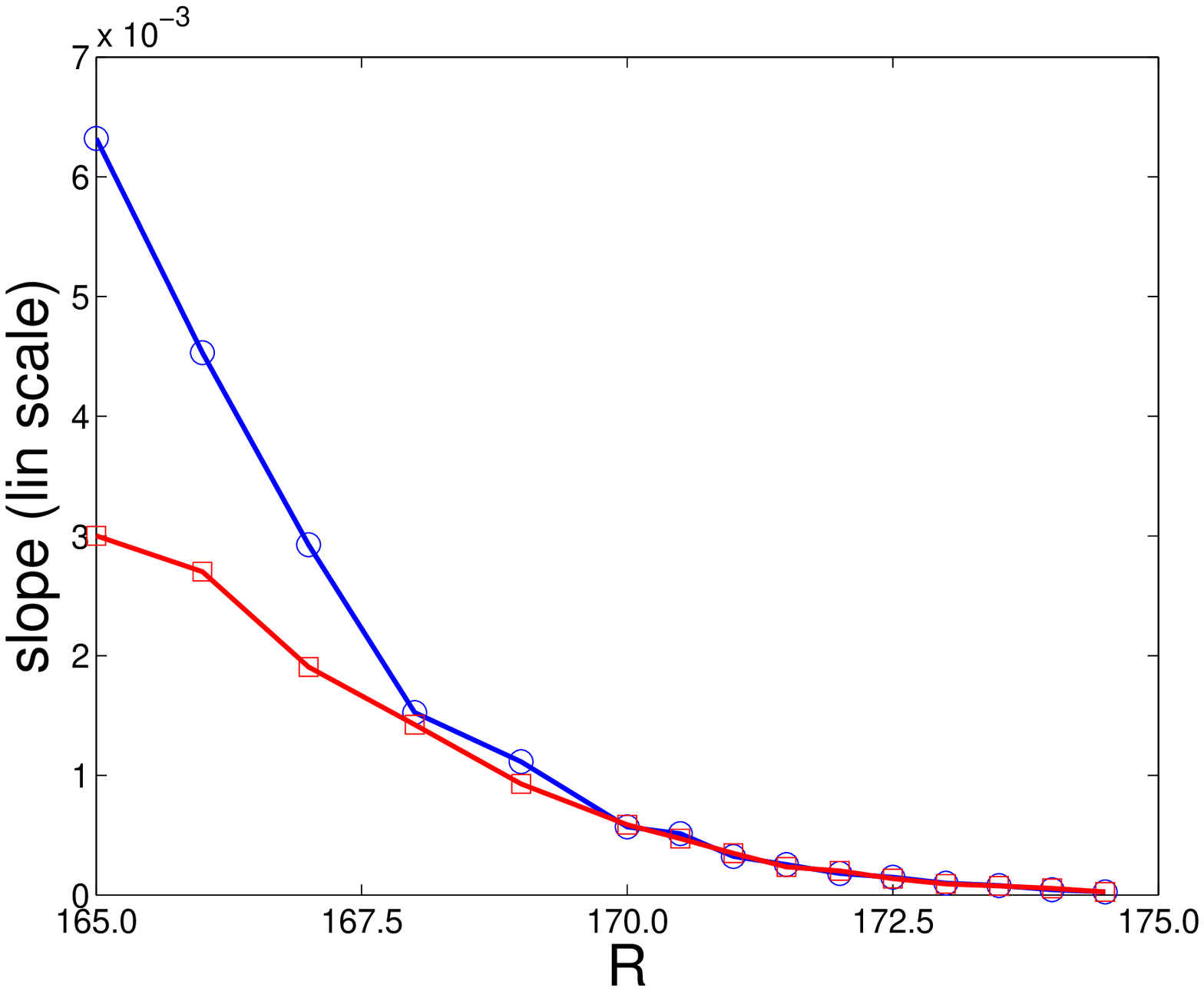}\hfill
\includegraphics[width=0.49\textwidth,clip]{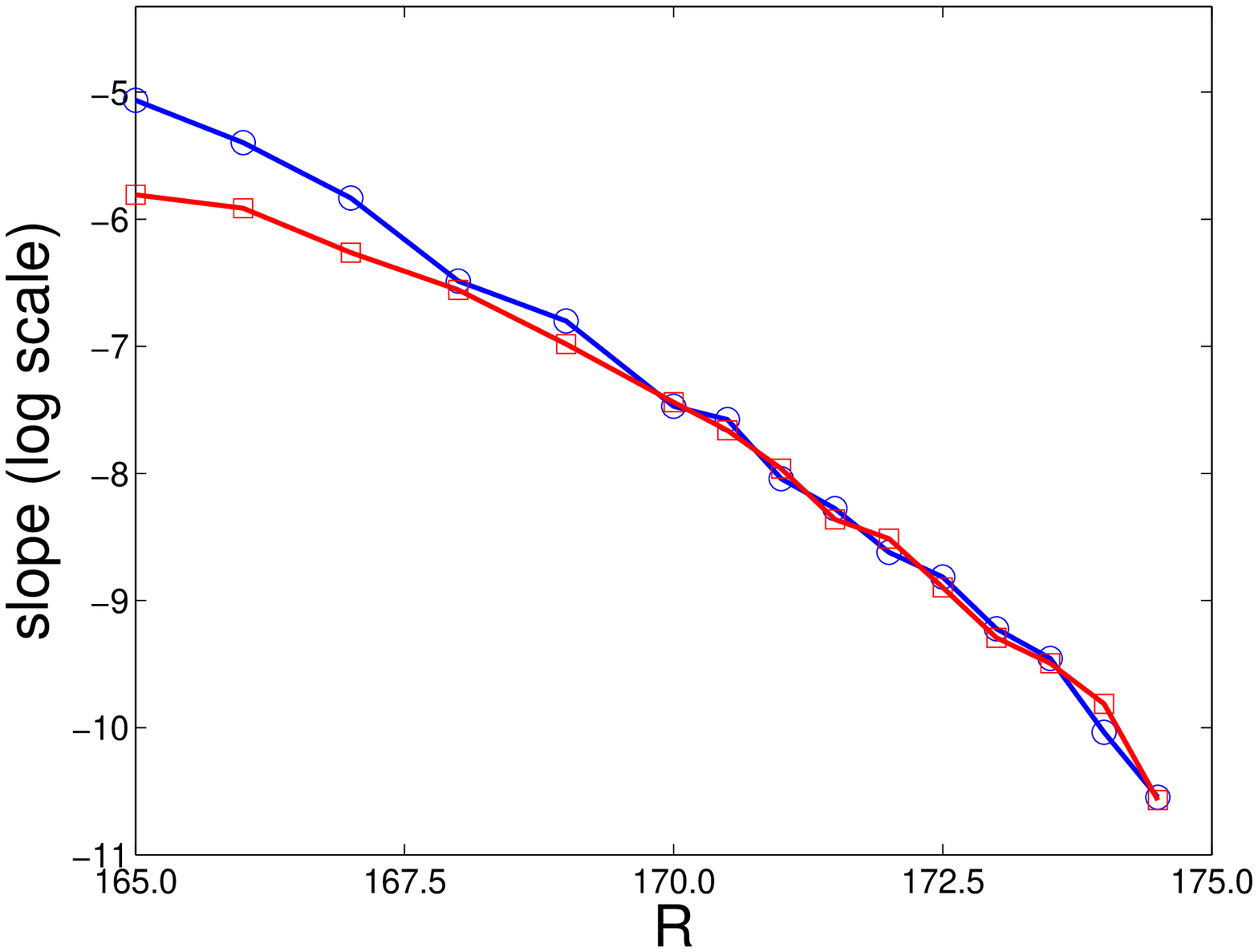}
\end{center}
\caption{Slope of histogram tails in lin-log scale (Fig.~\ref{ftv}) as
  a function of $R$. Top: lin-lin. Bottom: lin-(natural)log scale. Open dots:
  slope of histogram tails in semi-log plot; open squares: inverse mean
  length of transients computed as indicated in the text.
\label{ftw}}
\end{figure}

The slopes of the exponential tails were determined by fitting them
with straight lines. The result is plotted in Figure~\ref{ftw} as open
dots, either in linear scale (top) or semi-log scale (bottom). If the
distribution were exponential from the start, up to some offset value
$\tau_{\rm min}$, the slope would be given by
$1/\langle \tau-\tau_{\rm min}\rangle$ where $\langle \dots\rangle$
is the arithmetic mean over the observed transient lengths $\tau$.
Corresponding  results are plotted as open squares in
Fig.~\ref{ftw}. Both estimates compare well with each other down to
$R\sim169$ below which they diverge. In fact, histograms for $R<170$
were not presented in Fig.~\ref{ftv} in order not to overcrowd it with
curves that would have needed another scale to make them
discernable. These histograms display a shoulder corresponding to an
accumulation of short transients that biases the computation of the
slope from the inverse mean and explains the discrepancy between the
two estimates. Conversely this assures us that the fit to
an exponential is indeed satisfactory for $R\ge170$.

It is, however, hard to put error bars on our results because we have a
finite number of transients at our disposal, usually between 100 and 200
except for $R=174$ and $174.5$ for which we have only about 30
transients. To understand the origin of the difficulties, one must
consider how the histograms evolve as the number of experiments
increases. For example, the presence of a small number of measurements
that seem atypical of the statistical tail implies an increase of the
number of experiments by a large factor to rub out their effect. All
the values plotted in Figure~\ref{ftw} are what we consider to be our {\it
  best estimates\/} based on careful inspection of individual
histograms and comparisons of statistics obtained by including or
excluding transient lengths that might be judged atypical of the
overall decay behavior. In general all results were
consistent to within less than 10\%, especially at the largest values of
$R$ considered.

\begin{figure}
\begin{center}
\includegraphics[width=0.49\textwidth,clip]{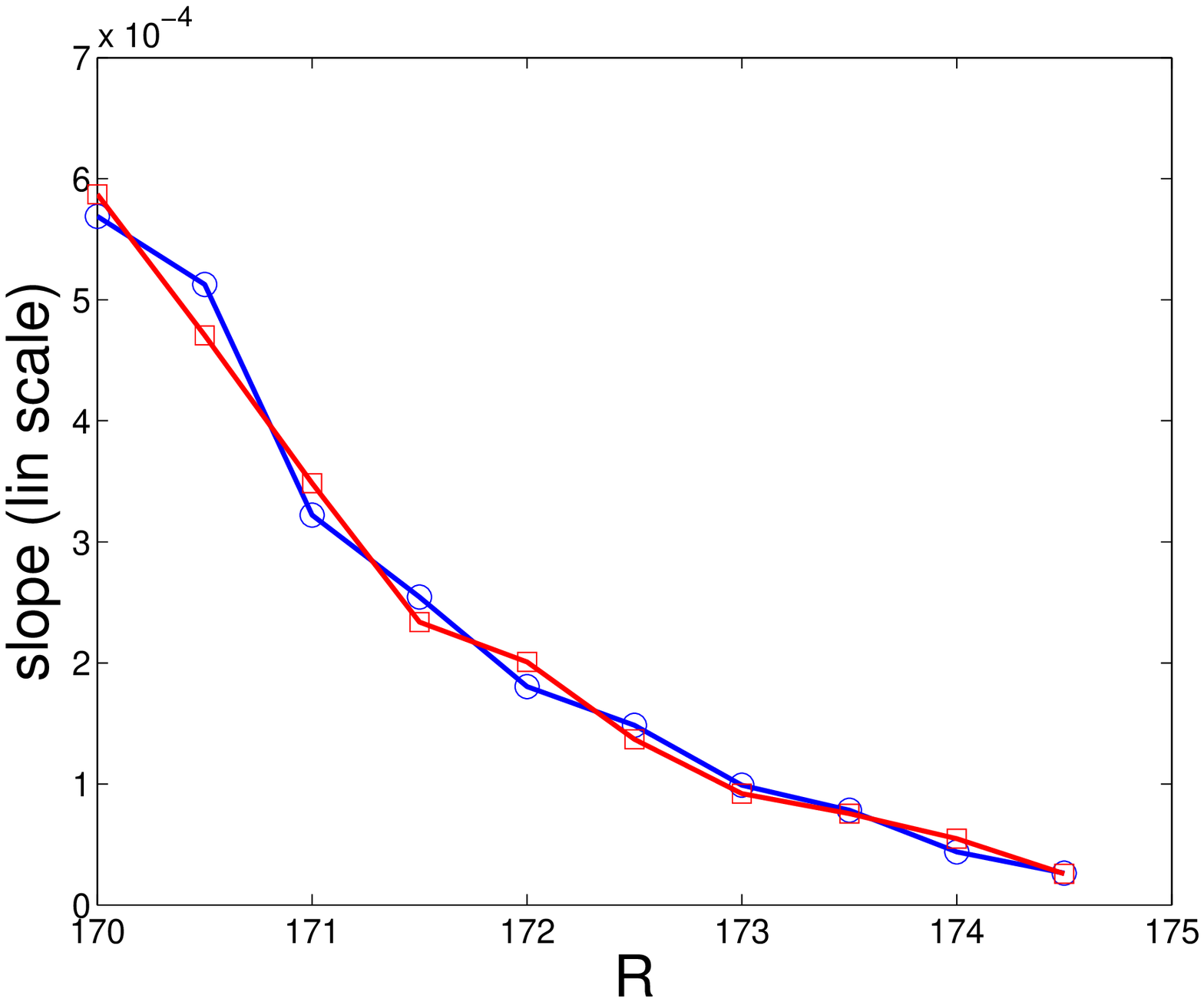}\hfill
\includegraphics[width=0.49\textwidth,clip]{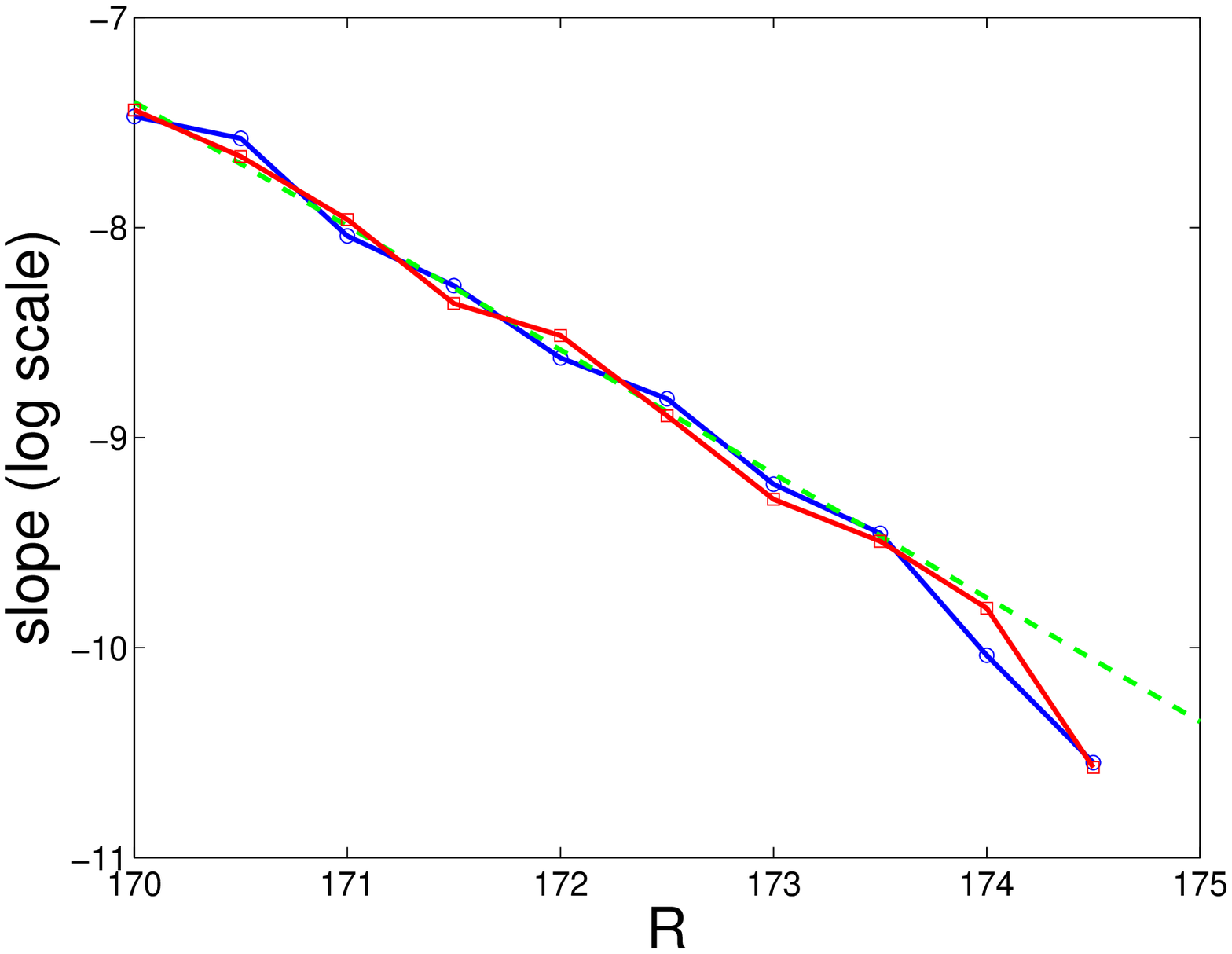}
\end{center}
\caption{Slope histogram tails in lin-log scale as a function of
  $R$. Enlargement of Fig.~\ref{ftw}. Top: lin-lin scale. Bottom:
  lin-(natural)log scale; the added dashed line is a guide for the
  eyes, showing the deviation of the top highest points from the
  overall exponential behavior. Symbols as in Fig.~\ref{ftw}.
\label{ftwb}}
\end{figure}

This being said, the zoom on the variation of the slope as a function of
$R$ displayed in Figure~\ref{ftwb} confirms the exponential decrease
of the terminal slope of histograms, or
conversely an exponential increase of the characteristic decay
time. However, the values corresponding to the points at $R=174$ and
$174.5$ seem somewhat too small for them to stay aligned with the other
points, as suggested by the dashed line in Figure~\ref{ftwb}
(bottom). This line is obtained from a linear fit of the logarithms of
the slopes measured for $R\in[170,173.5]$, which we consider as our
most reliable estimates. The deviation can have two explanations:\\
--\ either, as discussed earlier, it is an artifact of our estimation
procedure with an insufficient number of points but it turns out that the
corresponding distributions were surprisingly regular and the
slope estimation unambiguous,\\
--\ or it is a real effect indicating that the characteristic
time increases faster than expected, i.e. diverges for some finite
$R_{\rm g}$ that should be close to 175 (for which the
turbulent attractor was found to be stable).

For $R=174$ and $174.5$,  an over-representation of long transients
among our 30 or so samples would indeed explain the observed
under-evaluation of the slope, which might be the trace of a genuine
(turbulent) attractor for slightly larger values of $R$. Note also that
the difference between the estimates obtained by slope fitting and by inverse 
means at $R=174$ ---quite anomalous in view of the general agreement
between the two estimates when $R\ge170$--- is entirely accounted for by
the accidental absence of a sufficiently short transient
$\tau_{\rm min}$.
Accordingly, we do not exclude the possibility of a crossover between
the exponential regime typical of transients tentatively attributed to
a chaotic saddle and a real divergence at some well defined
$R_{\rm g}$ above which a turbulent attractor does exist. This divergence
would be typical of a crisis converting the chaotic repellor into
an attractor. The anomalously long transients observed for $R=174$ and
$174.5$ would then be representative of the behavior just below the
crisis point.

Finite size effects, still sizable at $\mathcal
D=32\times32$, will make it difficult to decide what is the actual
behavior but we believe that considering larger systems, which is
presently in progress, will help us to obtain a clearer picture of the
transition. The next section is thus devoted to an illustration of a
transient in a system of intermediate size with the purpose of showing
that one cannot avoid the spatio-temporal feature of the
dynamics in the transitional range of  Reynolds numbers.

\begin{figure}
\begin{center}
\includegraphics[width=0.66\textwidth,clip]{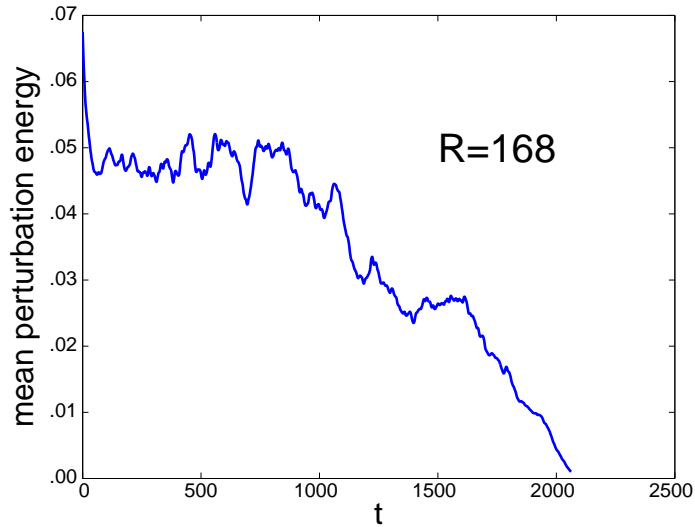}
\end{center}
\caption{Mean perturbation energy as a function of time during a
  transient in a quench experiment at $R=R_{\rm f}=168$ starting from
  $R_{\rm i}=200$ for $L_x=128$, $L_z=64$.\label{diag}}
\end{figure}

\begin{figure}
\begin{center}
\includegraphics[width=0.48\textwidth,clip]{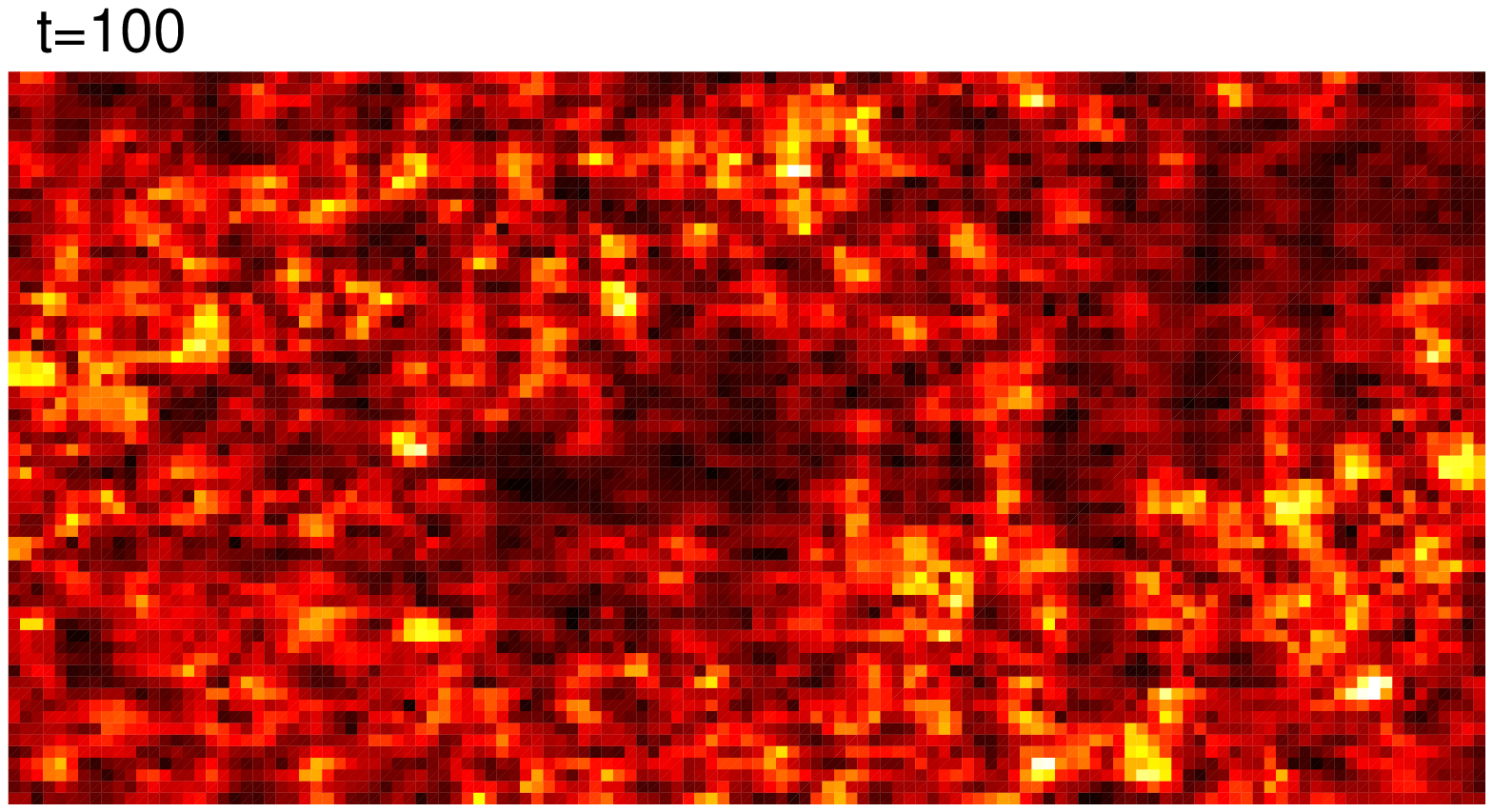}\hfill
\includegraphics[width=0.48\textwidth,clip]{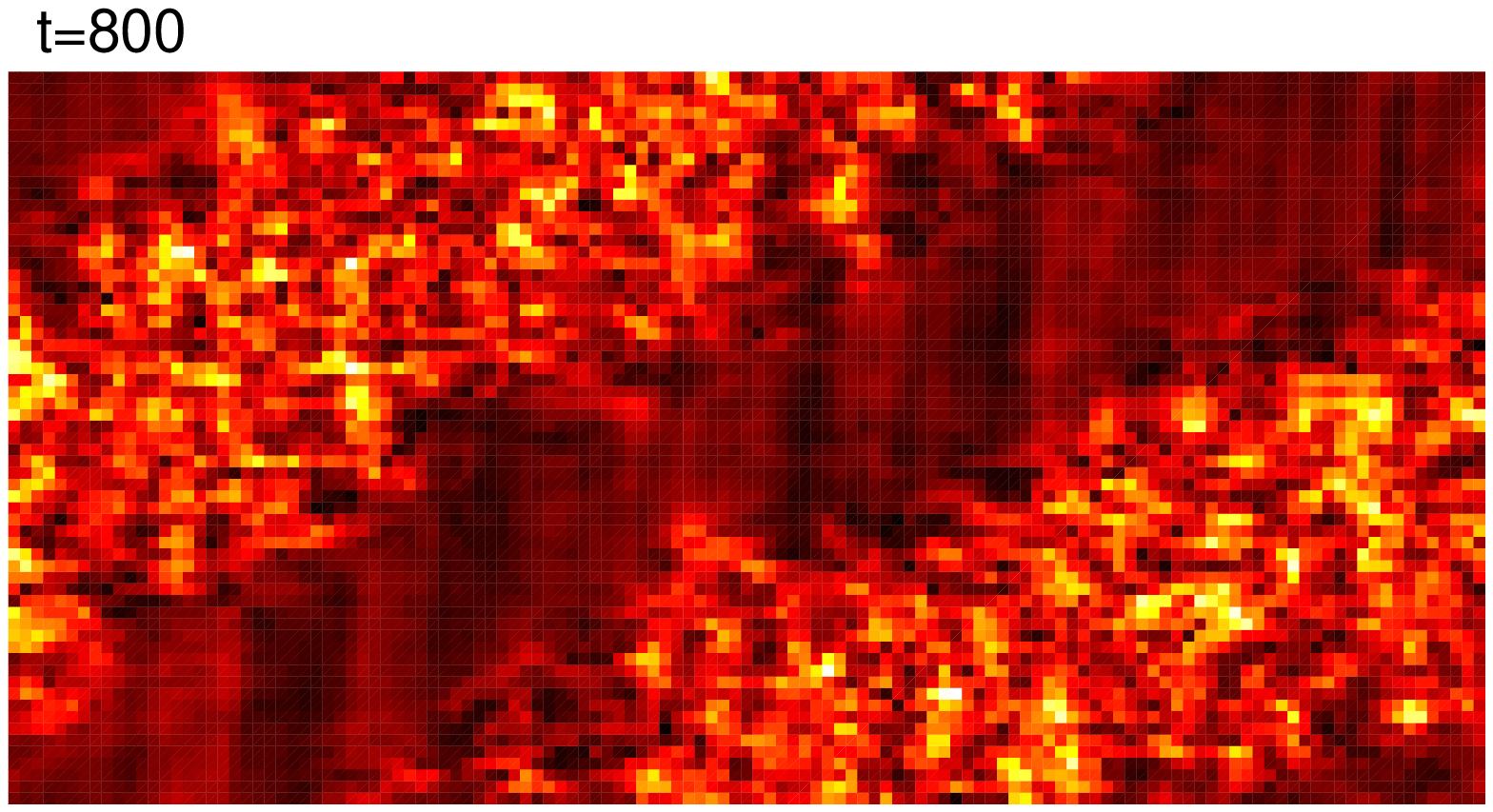}

\includegraphics[width=0.48\textwidth,clip]{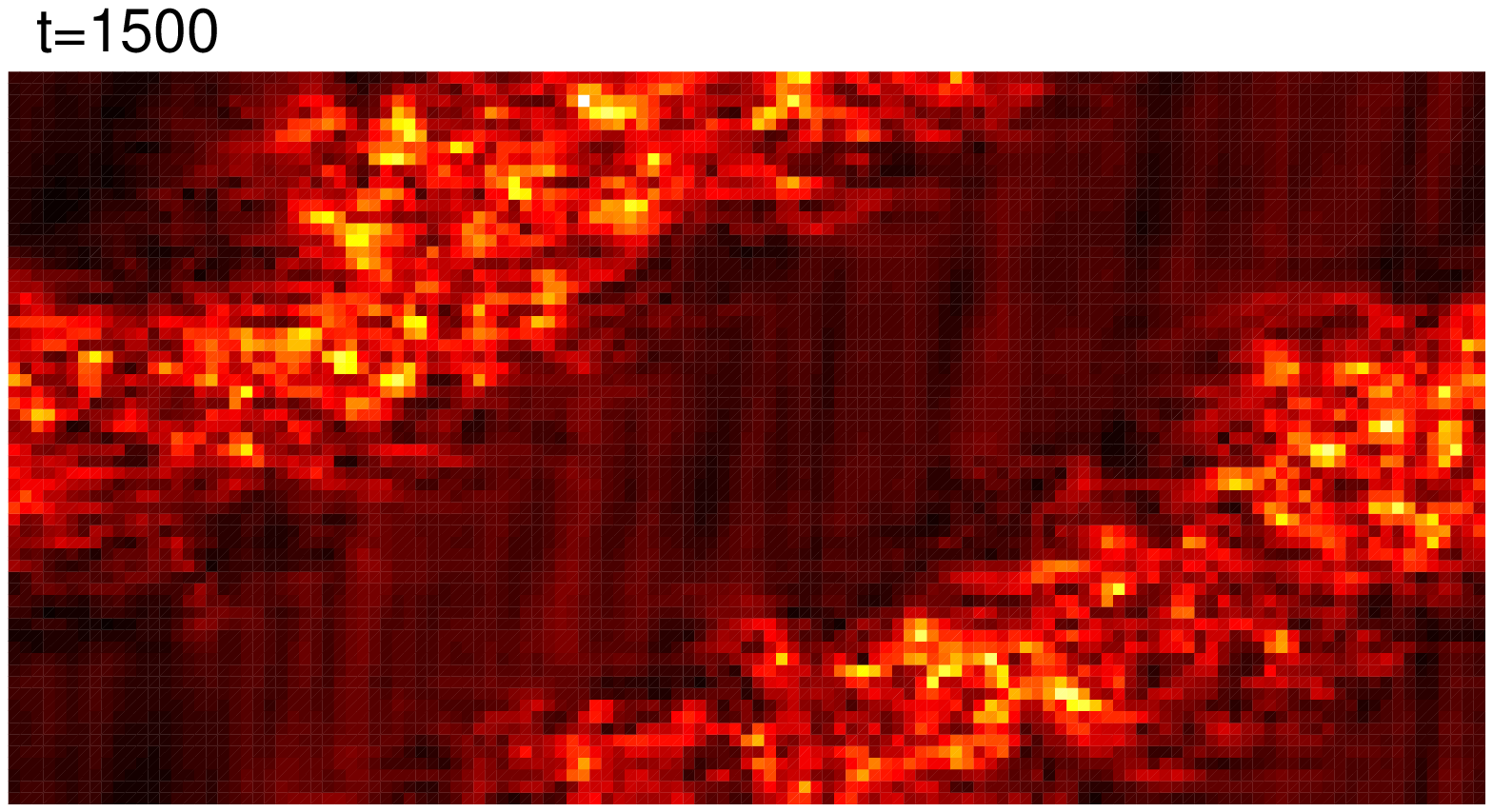}\hfill
\includegraphics[width=0.48\textwidth,clip]{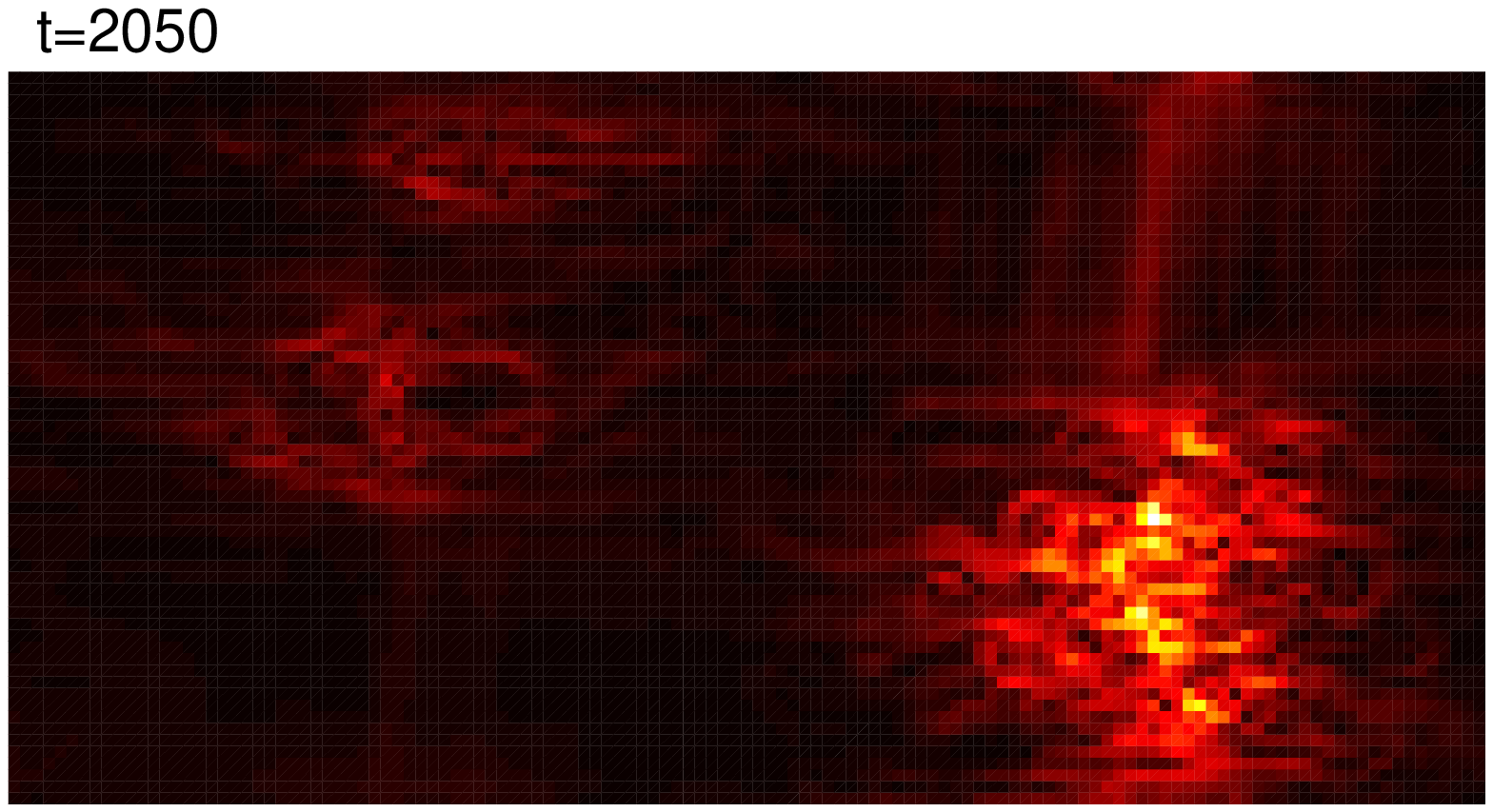}
\end{center}
\vspace*{-4ex}

\caption{Four snapshots taken during a transient in a domain $\mathcal
  D=128\times64$. Dark/deep red (clear/light yellow) regions 
  correspond to small (large) perturbation speed [colored version on
  line]. One point out of 4
  in each direction is represented, hence $\Delta x=\Delta z=1$. The
  horizontal axis corresponds to the streamwise direction. The
  contribution of streaks to the perturbation is clearly visible at
  the laminar/turbulent frontier.\label{snsh}}
\end{figure}

\subsection{Illustration of decay during a transient at $R=168$\label{S-tr}}

In this subsection we consider a typical transient obtained for
$R=168$ with $L_x=128$, $L_z=64$, starting from $R=R_{\rm i}=200$.
Figure~\ref{diag} displays the evolution of the mean perturbation
energy as a function of time throughout the transient. The general
aspect of this curve is similar to
those in Figure~\ref{ftu}, with a distinct plateau followed by
an irreversible decay. Here, the transient is not very long since
$R_{\rm f}$ is not close enough to $R_{\rm g}$. Snapshots of the local
perturbation velocity amplitude $\bar u(x,z,t)$ are given in figure
\ref{snsh}. This quantity is defined as 
$\bar u(x,z,t)=\sqrt{2(E_0(x,z,t)+E_1(x,z,t))}$, where
$E_0=\frac12(U_0^2+W_0^2)$ is the local kinetic energy contained in the
streaks and $E_1=\frac12(U_1^2+V_1^2+W_1^2)$ the kinetic energy
contained in the remaining part of the departure from the laminar
profile, including the contribution of streamwise vortices. This
definition is reminiscent of that of a turbulent velocity
intensity except that the latter is defined with respect to the mean
flow. Here, this would have led us to subtract $\overline{U}_1$, but
would no longer have made sense for the spatially inhomogeneous
turbulent regime during decay, as illustrated below.

Figure~\ref{snsh} presents several snapshots of the perturbation flow
pattern at different key instants. First for $t=100$, just after the
quench, turbulence initially prepared at $R_{\rm i}=200$ gets
stabilized at the level corresponding to $R_{\rm f}=168$. Though
it is relatively uniform, one can already detect some large scale
inhomogeneity. Later, this inhomogeneity increases up to a point when
oblique bands are conspicuous, as in the second image at
$t=800$ while the level of perturbation energy has remained
essentially the same. Subsequently, the turbulent region becomes narrower
and the perturbation energy decreases accordingly. At $t=1500$,
$E_{\rm t}$ is about the half of the initial energy content. The
regular regression of the band ends when it breaks into several parts
forming smaller patches that, in turn, decay as seen here for $t=2000$.
The regular regression of the width of the turbulent bands explains
the roughly linear decrease of the mean perturbation energy for
$T>1000$ in Fig.~\ref{diag} assuming that the front between
the turbulent and laminar domains moves at constant speed.

\begin{figure}
\begin{center}
\includegraphics[width=0.66\textwidth,clip]{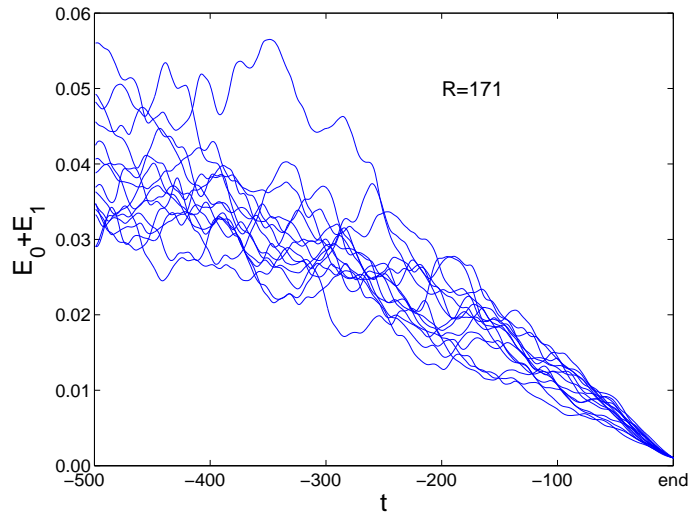}
\end{center}

\caption{Variation of the total perturbation energy during the end of
  15 of the transients in Fig.~\ref{ftu}.\label{endtr}}
\end{figure}

In fact, a similar behavior was probably exhibited by transients
already for $\mathcal D=32\times32$ though the picture in physical
space is less clear. Indirect evidence can be obtained by zooming 
on the last 500 times units of each transient. Figure~\ref{endtr}
displays a superposition of 15 of them for $R=171$ taken from the
series used for Fig.~\ref{ftu}. Statistically speaking, they all follow
the same roughly linear decay, from the end of the plateau to
the cut-off corresponding to the beginning of the final exponential
viscous damping (not represented). Accordingly, and comparing to
what happens for $\mathcal D=128\times64$, it seems that we
have a nucleation problem with a probability distribution
for laminar germs of different sizes so that when a wide enough germ
appears, the energy plateau ends and the turbulent domain recedes
regularly. (At this stage, it could be noticed that uniformly
subtracting about 500 time units from the transient lengths would not
change the overall statistical behavior.) The intensity of the
mechanism sustaining turbulence, a function of $R$, is certainly
essential in controlling this probability. Such a mechanism is
sometimes studied using low dimensional dynamical systems modeling
\cite{Wa97,EM99,SMH05}. Accordingly,
a way to reconcile this picture with that of 
chaotic saddles could be recovered but it is still not clear how to
reintroduce physical space and correlative size effects. At any rate,
considering a larger system will certainly shed light on this
nucleation probability problem. 

It is worth to mention that, during the decay and even at rather late
stages, the energy  remains much concentrated in the turbulent
patches. It was empirically found that setting the cut-off between
turbulent and laminar flow at $\bar u=0.5\bar u_{\rm ref}$, where
$\bar u_{\rm ref}$ is the mean of $\bar u$ in the starting reference state
at $t=100$, well discriminates the two regimes from each other.
For example, at $t=2000$, one finds that about 70\% of the remaining
energy is concentrated in 10\% of the total surface. Going backwards in
time, one gets that, at $t=1500$, the turbulent band occupies about
48\% of the surface and contains around 90\% of the energy, and still
earlier, that for $t=800$, the turbulent fraction is about 87\% and
contains 98\% of the energy.

To conclude this section, it should be remarked that the oblique
band observed in Fig.~\ref{snsh} has no connection 
with the oblique turbulent bands experimentally observed in the
upper transitional range \cite{Petal02} and numerically reproduced
by Barkley and Tuckerman \cite{BT05}. Here, its presence is clearly
linked to the periodic boundary used and its 
development directly follows from the oblique ovoid shape of the
inhomogeneity already present at the very beginning of the quench
experiment. Depending on the size and average orientation of this
inhomogeneity, the ultimate shape of the turbulent domain is either a
longitudinal, a transverse, or an oblique band. In view of the size of
modulations at $t=100$ (Fig.~\ref{snsh}, top), we think that a system
at least four times as large as that used for this illustration
should be considered in order to have a chance to 
explore several inhomogeneity patterns before decay. This 
is what motivates the simulations over a domain $\mathcal D=256\times128$
presently in progress and alluded to above.

\section{Conclusion\label{S-CONC}}

In this paper we have considered the low-$R$ range of transitional
pCf. On the one hand, modeling this situation in terms of low
dimensional deterministic dynamical systems is limited due to the
inherently spatiotemporal character of the problem. For example, decay
from the sustained turbulent state 
close to the global stability threshold $R_{\rm g}$ is not well
described as a chaotic transient close to a crisis bifurcation 
point, even though exponential distributions of transient lifetimes
are observed near this point \cite{MFE04}, because this concept
introduced in nonlinear dynamics is not adapted to the account of
decay through regression of turbulent patches fluctuating in time and
space. On the other hand, the direct connection to statistical physics
suggested by Pomeau \cite{Po86} and  underlying the abstract
spatio-temporal intermittency approach \cite{CM95} may seem far-off
though it takes spatial extension into account in an analogous way.
Extending previous work by one of us \cite{ML00,Ma04}, the approach
developed here intends to bridge the gap between low and high
dimensional systems in a concrete way and with a semi-quantitative
ambition.

Except for the fact that it fell short of predicting 
reasonable values of the transitional Reynolds numbers, the early
stress-free attempt already contained many interesting features and
its deficiency could easily be attributed to the use of unrealistic
boundary conditions in the derivation of the model.
The first part of the present paper was thus devoted to the derivation
of a similar model, but appropriate to the physically realistic
no-slip boundaries \cite{GP71,Ma90,Ma83}. When truncated at the lowest
significant order (\S\ref{S-nsl}), this new model was seen to have
nonlinear and non-normal structures close to those of the
stress-free model but with slightly different coefficients when basis
functions used in the Galerkin expansion were appropriately normalized.
The major differences, discussed in \S\ref{S-COMP}, appeared to be
({\it i\/}) the presence of an additional damping of the drift flow
that was no longer cross-stream independent but had a more generic
parabolic profile, ({\it ii\/}) a strongly enhanced viscous
dissipation of the other flow components (associated to steeper
variations of the Galerkin basis functions close to the plates), and
({\it iii\/}) the existence of a non-trivial mean correction to the
base flow $\overline{U}_1$ already at lowest truncation order.
As could be expected from the structural similarity of the models,
the same qualitative behavior was obtained but a wide step in the
right direction was made since the study presented in the
second part of the paper showed that transitional Reynolds numbers of
interest were now less than a factor of two off the experimental
range, instead of a factor for ten with the stress-free model.

Experiments reported in~\cite{Betal98a,KNN05} clearly showed that, in the
low-$R$ part of the bifurcation diagram, the dynamics is dominated by
processes involving flow structures that are coherent over the full
gap between the plates and this is certainly the reason why the
results obtained from the model are reasonably satisfactory. Increased
dissipation associated to more resolved cross-stream dependence is
expected to improve the agreement at the price of much more analytical
and computational work already around $R_{\rm g}$.

On the other hand, increasing the order of the model would certainly
be necessary much beyond $R_{\rm g}$, in particular in the range where
the transition `oblique turbulent bands $\leftrightarrow$ featureless
turbulence' takes place \cite{BT05}. This is however not advisable
since the model would then get more and more cumbersome explicit nonlinear
terms from the Galerkin expansion which, in turn, would be truncated at
increasingly insufficient orders, while it is clearly more economical
to use a high precision pseudo-spectral scheme in the cross-stream
direction.

Besides the observation that fully developed spatiotemporal chaos is
extensive (\S\ref{S-EXT}), the main concrete result presented here 
relates to the determination of a possible global stability threshold
$R_{\rm g}$ and the statistical behavior of transients close to
$R_{\rm g}$ for systems of moderate size. 
As in laboratory experiments, a discontinuous transition was
observed (\S\ref{S-SUB}, Fig.~\ref{ftt}). Turbulent transients
accounting for the relaxation of turbulence during quench experiments
were shown to exhibit exponentially distributed lifetimes
(\S\ref{S-GST}, Fig.~\ref{ftv}). Beyond 
$R\simeq175$ the turbulent state was observed over time intervals that
could seemingly be arbitrarily long, hence suggesting $R_{\rm g}=175$.
For $R<175$, the slopes of the exponential tails of the lifetime
distributions were seen to approach zero exponentially as $R$
increased, with a possible crossover to a diverging behavior close to
the supposed $R_{\rm g}$, in spite of post-processing uncertainties
(Fig.~\ref{ftwb}).

Extending this conclusion to the case of transitional pipe
Poiseuille flow \cite{Hetal06,PM06,WK07} needs additional
investigation before tentatively establishing any connection.
Such a connection, which should be taken with care, has been made
from a reanalysis of the pCf results that also leads to indefinite
exponential decay with $R$ of the slopes of histograms' tails. In our
opinion, in the Couette case available experimental data were however
obtained for values of $R$ somewhat too far from the hypothesized
$R_{\rm g}\simeq325$ (determined from several independent
ways) for supporting the conjecture of an underlying chaotic saddle
typical of temporal chaos, especially in view of experimental pictures
\cite{Betal98b}. On the contrary, in close correspondence with these
pictures, the illustration given in \S\ref{S-tr} clearly points to  
spatio-temporal chaos as explaining the transient behavior, with the
statistical possibility of a local breakdown of the turbulent state,
further extending to the whole system. This behavior is reminiscent of
what happens at a thermodynamic phase change \cite{Po86}. The size of
the system for which these preliminary results have been obtained, 
$\mathcal D=32\times32$, though situating it apparently inside the
extensive regime, is clearly insufficient to probe any
thermodynamic-like property of the transition, and {\it a fortiori\/}
any issue related to universality in the sense of critical phenomena
in statistical physics. We expect that the study currently in
progress for $\mathcal D=256\times128$, a size unreachable by direct
numerical simulations, will bring us interesting results.

Since it represents a simplified version of the Navier--Stokes
equations at least for $R$ not too large, our model can serve to
study other questions posed by the transition to turbulence in globally
sub-critical flows. For example, whereas it is clear how streamwise
vortices generate streaks through the {\it lift up\/} mechanism
introduced long ago, instability mechanisms and some processes
involved in the regeneration of vortices are still unclear in spite of
recent progress \cite{MK05}. The very coexistence of laminar and
turbulent flow implied by global sub-criticality, the
mechanisms by which one of the phases gains against the other,
i.e. how spots grow ({\it laminar $\to$ turbulent\/} transition) or how the
turbulent state recedes ({\it  turbulent $\to$ laminar\/} transition), how
asymmetric these problems are, what is the role of the base flow
modification inside a turbulent domain (local counter-part to
\S\ref{S-MF} in the presence of large scale modulations), are
questions that we intend to examine, with the 
stimulating perspective of going beyond the case of pCf and
considering other wall flows of less academic interest but
experiencing a similar transition to turbulence.

\section*{Appendix: Equations used in simulations}
The equations for $\Psi_0$, $\Psi_1$, and $\Phi_1$ that have been
effectively implemented read:
\BAN
&&\left[\D_t-R^{-1}\left(\Delta-\gamma_0\right)\right]\Psi_0\\
&&\quad =\mbox{}\Delta^{-1}(\D_z N_{U_0}-\D_x N_{W_0})
+a_1\left(\sfrac32\D_z\Phi_1-\D_x\Psi_1\right)\,,\\
&&\left[\D_t-R^{-1}\left(\Delta-\gamma_1\right)\right]\Psi_1\\
&&\quad=\mbox{}\Delta^{-1}(\D_z N_{U_1}-\D_x N_{W_1})-a_1\D_x\Psi_0\,,\\
&&\left[\D_t
-R^{-1}\left(\Delta-\beta^2\right)^{-1}
\left(\Delta^2-2\beta^2\Delta+\gamma_1\beta^2\right)\right]
\Phi_1\\
&&\quad=\mbox{}\left(\Delta-\beta^2\right)^{-1}
\left[\Delta^{-1}\beta^2(\D_x N_{U_1}+\D_z N_{W_1})-\beta N_{V_1}\right]\,,
\EAN
where $\beta=\beta^{\rm ns}=\sqrt3$ everywhere and the expressions of
the advection terms have been rewritten in the energy conserving form
mentioned in the text:
\BAN
N_{U_0}&=& M_{U_0} + \partial_x E_{\rm a}\,,\\
N_{W_0}&=& M_{W_0} + \partial_z E_{\rm a}\,,\\
N_{U_1}&=& M_{U_1} + \partial_x E_{\rm b}\,,\\
N_{W_1}&=& M_{W_1} + \partial_z E_{\rm b}\,,\\
N_{V_1}&=& M_{V_1} + \beta E_{\rm b}\,,
\EAN
where the ``weighted'' energies read:
\BAN
E_{\rm a}&=& \sfrac12 \alpha_1(U_0^2+W_0^2)+
\sfrac12\alpha_2(U_1^2+W_1^2)+\sfrac12\alpha_3 V_1^2\,,\\
E_{\rm b}&=&\alpha_2(U_0U_1+W_0W_1)\,,
\EAN
and
\BAN
M_{U_0}&=& V_1(\alpha_2\beta' U_1-\alpha_3\partial_xV_1)
+\alpha_1W_0(\partial_z U_0-\partial_x W_0)\\
&&\mbox{ }+ \alpha_2 W_1(\partial_zU_1-\partial_x W_1)\,,\\
M_{W_0}&=&\alpha_1U_0(\partial_x W_0-\partial_z U_0)
+\alpha_2 U_1(\partial_xW_1-\partial_z U_1)\\
&&\mbox{ }+ V_1( \alpha_2\beta' W_1-\alpha_3\partial_zV_1)\,,\\
M_{U_1}&=& \alpha_2[W_0(\partial_z U_1-\partial_x W_1)
+W_1(\partial_z U_0-\partial_x W_0)]\\
&&\mbox{ }- \alpha_2\beta'' V_1U_0 \,,\\
M_{W_1}&=&\alpha_2[U_0(\partial_x W_1-\partial_z U_1)
+U_1(\partial_x W_0-\partial_z U_0)]\\
&&\mbox{ }- \alpha_2\beta'' V_1W_0\,,\\
M_{V_1}&=& U_0(\alpha_3\partial_xV_1-\alpha_2\beta U_1)\\
&&\mbox{ } + W_0(\alpha_3\partial_zV_1-\alpha_2\beta W_1)\,.
\EAN
Velocity components are given by (\ref{E-Psi0}--\ref{E-Phik}).

The evolution is computed in Fourier space by a numerical scheme that
treats linear terms exactly and nonlinear terms by an Adams--Bashforth
formula. Formally writing the evolution problem as
$$
\sfrac{\rm d}{{\rm d}t} X=\mathcal L X + \mathcal N(X)
$$
it can be seen that, in order to keep second order accuracy, we
have to compute
\BAN
&&X(t+\delta t)=\exp(\mathcal L\delta t) \times\\
&&\>\left[X(t)+\delta t\big(1.5\,\mathcal N(X(t))
-0.5\exp(\mathcal L\delta t)\mathcal N(X(t-\delta t))\big)\right]\!.
\EAN
When integrating the equations, it can also be verified that their
r.h.s. cancel exactly at wave-vector $(k_x,k_z)=(0,0)$ so that taking
the inverse of $\Delta$ is a legitimate operation. As stated in the
main text, average velocity components must be treated separately.
Corresponding formulas for ($\overline U_1,\overline U_0)$ are obtained
from (\ref{E-U1bar},\ref{E-U0bar}) and those for 
and $(\overline W_1,\overline W_0)$ by two similar equations with
$U$ replaced by $W$.
\bigskip

\noindent {\large\bfseries\sffamily Acknowledgments}
\medskip

\noindent The authors acknowledge discussions with J.-M. Chomaz, C. Cossu,
and P. Huerre at LadHyX, B. Eckhardt and J. Schumacher at
Phillipps Universit\"at, Marburg.
P.M. specially thanks S.~Bottin, O.~Dauchot, F.~Daviaud, and
A.~Prigent at Saclay for their past and present collaboration. L.~Tuckerman and
D.~Barkley are thanked for interesting discussions and the
communication of their own papers related to the problem treated here.
The two anonymous referees are also deeply
acknowledged for their thoughtful and constructive remarks that helped
us to improve the manuscript.

Computations at $L_x=128$, $L_z=64$ were performed at IDRIS (Orsay)
under project \#~61462.

\end{document}